\documentclass[prd,superscriptaddress,twocolumn,showpacs]{revtex4}
\usepackage{bm} 
\usepackage{graphicx}
\newcommand{\be}{\begin{equation}}
\newcommand{\ee}{\end{equation}}
\newcommand{\bea}{\begin{eqnarray}}
\newcommand{\eea}{\end{eqnarray}}
\begin{document}
\title{Dijet production as a centrality trigger for $pp$ collisions 
at CERN LHC}
\author{L.~Frankfurt}
\affiliation{School of Physics and Astronomy, Tel Aviv University, 
Tel Aviv, Israel}
\author{M.~Strikman}
\affiliation{Department of Physics, Pennsylvania State University,
University Park, PA 16802, USA}
\author{C.~Weiss}
\affiliation{Institut f\"ur Theoretische Physik,
Universit\"at Regensburg, D--93053 Regensburg, Germany}
\begin{abstract}
We demonstrate that a trigger on hard dijet production at small rapidities 
allows to establish a quantitative distinction between central and 
peripheral collisions in $\bar p p$ and $pp$ collisions at Tevatron and 
LHC energies. Such a trigger strongly reduces the effective impact 
parameters as compared to minimum bias events. This happens because
the transverse spatial distribution of hard partons ($x\agt 10^{-2}$)
in the proton is considerably narrower than that of soft partons,
whose collisions dominate the total cross section. In the central 
collisions selected by the trigger, most of the partons with 
$x\agt 10^{-2}$ interact with a gluon field whose strength 
rapidly increases with energy. At LHC (and to some extent already at 
Tevatron) energies the strength of this interaction approaches 
the unitarity (``black--body'') limit. This leads to 
specific modifications of the final state, 
such as a higher probability of multijet events
at small rapidities, a strong increase of the transverse momenta 
and depletion of the longitudinal momenta at large rapidities,
and the appearance of long--range correlations in rapidity between 
the forward/backward fragmentation regions. The same pattern is expected 
for events with production of new heavy particles (Higgs, SUSY). 
Studies of these phenomena would be feasible with the CMS--TOTEM detector 
setup, and would have considerable impact on the exploration of 
the physics of strong gluon fields in QCD, as well as the search 
for new particles at LHC.
\end{abstract}
\maketitle
\section{Introduction}
\label{sec_introduction}
\begin{figure}[b]
\includegraphics[width=8cm,height=5.3cm]{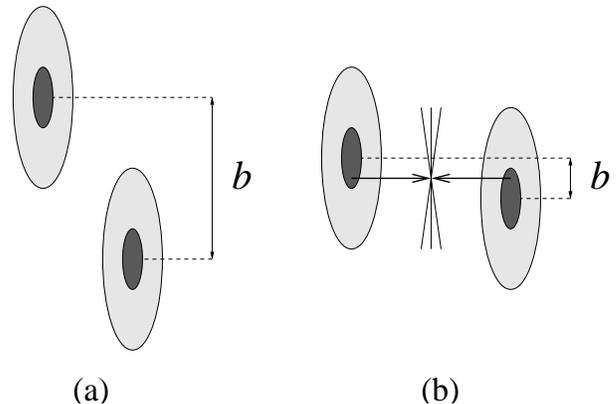}
\caption[]{Schematic illustration of the two classes of $pp$ collisions
at high energies. The transverse spatial distributions of the hard partons
($x \geq 10^{-2}$) is indicated by the dark shaded disks, those of the
soft partons ($x \ll 10^{-2}$) by the light shaded disks;
$b$ denotes the impact parameter of the $pp$ collision. 
(a) At large $b$ no overlap between hard partons occurs.
(b) At small $b$ the distributions of hard partons overlap, 
leading to production of hard dijets (and, possibly, heavy particles).}
\label{fig_coll}
\end{figure}

The differences between peripheral and central 
collisions play a crucial role in the physics of heavy ion collisions. 
In $pp$ collisions, a similar distinction can be made at the energies 
of the LHC and the Tevatron. This is possible because of the
appearance of two separate transverse distance scales at high energies.
On one hand, as predicted by Gribov \cite{Gribov:fm}, the essential 
impact parameters in hadron--hadron collisions increase with the 
energy. This has been observed \textit{e.g.}\ in numerous 
experiments in elastic $pp$ scattering, see Ref.~\cite{Block:1984ru} 
for a review. On the other hand, the transverse spatial distributions of 
hard partons (with finite light-cone fraction, $x$) in the colliding 
nucleons is only a weak function of $x$. For the gluon distribution
this has been verified experimentally in studies of the $t$--dependence 
of photoproduction of heavy quarkonia off the 
nucleon \cite{Frankfurt:2002ka}.
The two scales allow us to classify $pp$ collisions at collider energies.
A schematic illustration of this idea is given in Fig.~\ref{fig_coll}.
In collisions with large impact parameters there will be essentially 
no overlap between the hard partons (Fig.~\ref{fig_coll}a). Only partons 
with $x\ll 10^{-2}$ will overlap with significant probability. 
These peripheral events constitute a significant (in fact, dominant) 
part of the total inelastic cross section. The production of high 
$p_\perp$ jets, however, as well as of heavy particles, 
will be strongly suppressed. At small impact parameters, however, 
the distributions of hard partons in the two colliding nucleons 
will overlap, and the probability of hard interactions will be greatly 
enhanced (Fig.~\ref{fig_coll}b). This difference between the physics of 
soft and hard QCD processes (\textit{i.e.}, 
with $x_1, x_2$ of the colliding partons
$\geq 10^{-2}$) gives us an opportunity to distinguish quantitatively 
between central and peripheral collisions at collider energies. 

Specifically, we propose here to use the production of 
(one or more) hard dijets near zero rapidity as a ``centrality trigger'' 
for $\bar p p$ and $pp$ collisions at Tevatron and LHC energies. 
At the LHC, such a trigger could be implemented with any of the central 
detectors. It will lead to a significant enhancement of the production 
of hadrons at small rapidities and drastic changes of the pattern of 
forward production, which could be probed, for example, 
by the TOTEM detector in combination with the CMS detector.
Since the production of heavy particles, such as the Higgs boson or 
supersymmetric particles, is also greatly enhanced for central collisions, 
such a program could have considerable impact on the searches for new 
particles at the LHC.

\begin{figure}[t]
\includegraphics[width=8cm,height=2.6cm]{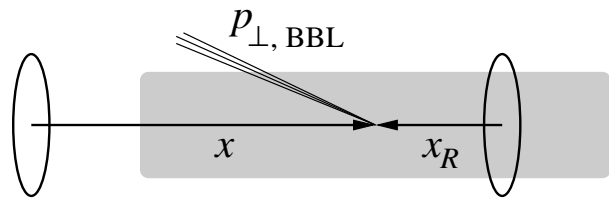}
\caption[]{Schematic illustration of the effect of the black-body
(unitarity) limit on hardon production in the forward/backward rapidity
region in central $pp$ collisions. A small--$x$ spectator parton 
(\textit{i.e.}, a parton not involved in the centrality trigger) from 
the left proton propagates through the strong gluon field
(indicated by the shaded area), acquiring a large transverse momentum, 
$p_{\perp, \text{BBL}} \gg \Lambda_{\text{QCD}}$. 
The small--$x$ parton is then resolved in a
collision with a large--$x_R$ parton from the right proton, resulting
in hadron production in the backward rapidity region.}
\label{fig_field}
\end{figure}

Another important application of the proposed ``centrality trigger''
is in the investigation of the physics of strong gluon fields in QCD. 
Numerous measurements of small--$x$ phenomena at HERA have confirmed
the fast increase of the gluon density in the proton predicted 
by perturbative QCD; for a review and references see 
Ref.~\cite{Abramowicz:1998ii}.
At LHC energies, and to some extent already at the Tevatron, 
the gluon density becomes so large that the interaction of high $p_\perp$ 
partons with the gluon ``medium'' becomes strong
and multiple scattering effects cannot be neglected, 
for a review and references see Ref.~\cite{Frankfurt:2000ty}. 
The new phenomenon one encounters here can be described as 
the breakdown of the Dokshitzer--Gribov--Lipatov--Altarelli--Parisi
(DGLAP) approximation caused by the approach to 
the unitarity limit(Black Body Limit, BBL), in which partons with 
virtualities below a certain value interact with the other proton with 
the maximal strength allowed by $s$-channel unitarity. More quantitatively,
a parton in one proton with longitudinal momentum fraction $x_R$ 
and virtuality $p_\perp^2$ resolves partons in the other proton with
\be
x \;\; = \;\; \frac{4 \, p_\perp^2}{x_R \, s} ,
\label{x_resolved}
\ee
where $s$ is the invariant energy squared of the $pp$ collision,
see Fig.~\ref{fig_field}. In particular, large--$x_R$ partons resolve 
small--$x$ partons in the other proton. At the LHC energy,
$\sqrt{s} = 14000 \, \text{GeV}$, for $p_\perp \sim 2\, \text{GeV}/c$ 
and $x_R \sim 10^{-2}$ one obtains $x \sim 10^{-5}$. Under this 
condition the interactions with gluons can approach the BBL,
see Ref.~\cite{Frankfurt:2000ty} and references therein. 
Present data on heavy quarkonium photoproduction indicate that
the gluon density at small $x$ is maximum in the transverse center 
of the proton \cite{Frankfurt:2002ka}. Since the distribution of 
large--$x_R$ partons is likewise concentrated at small transverse distances, 
it is evident that the chances of approaching the BBL are maximum for central
$pp$ collisions, which can be selected with the proposed 
``centrality trigger''.

The approach to the BBL in central $pp$ collisions leads to a 
significant change of the initial and final state interactions 
in the hard QCD processes. The interactions with the ``high density 
gluon medium''  suppress the spectrum of low transverse momentum partons 
and enhance the high momentum tail. Typically,
all partons which are not involved in the hard interactions at
the scale much higher than the BBL scale, $p_{\perp, \text{BBL}}$, 
receive transverse momenta $\sim p_{\perp, \text{BBL}}$ and experience a
significant momentum loss due to the gluon radiation
(a schematic illustration is provided in Fig.~\ref{fig_field}).
A forward would--be spectator component of the proton wave function
is ``pulverized'' completely, loosing its coherence.
As a result the number of particles with $x_F\geq 0.1$ in the proton
fragmentation region will strongly diminish, while the average
transverse momenta of these particles will grow to values comparable to
$p_{\perp, \text{BBL}}$. Much more energy will be released at 
small rapidities as compared to minimal bias events.

Although the phenomena discussed here are higher--twist corrections 
to the inclusive cross section and the transverse spectra of 
the production of sufficiently heavy particles (Higgs, SUSY), 
in the vicinity of the BBL
they will strongly modify the overall structure of the final states. 
In particular, they will change the pattern of radiation of 
moderate $p_\perp$ jets and the Sudakov form factors for dijet production.
Thus, the understanding of these phenomena is important also for 
effective searches for new physics at the LHC.

The basic idea of the proposed ``centrality trigger'' is that the
restriction to events with production of a hard dijet strongly reduces 
the effective impact parameters in high--energy $\bar p p$ and $pp$ 
collisions as compared to minimum bias events. One can further
narrow the distribution of impact parameters by requiring the presence 
of multiple dijets in the same event (this was first studied within the
framework of a multiple interaction Monte Carlo model in
Ref.~\cite{Sjostrand:1987su}). The actual reduction
which can be achieved in this way depends, to some extent, on possible 
spatial correlations of hard partons in the transverse plane. 
In this respect we make a surprising observation, namely that the data 
on double dijet production in $\bar p p$ collisions at the Tevatron obtained 
by the CDF Collaboration \cite{Abe:1997bp} indicate significant spatial 
correlations of partons in the transverse plane.

The paper is organized as following. In Section~\ref{sec_inelastic} 
we review the available information on the impact parameter
distribution in the generic inelastic $pp$ collisions.
In Section~\ref{sec_hard} we summarize our knowledge of the  
transverse spatial distribution of gluons in the 
nucleon and study its dependence on the resolution scale.
In Section~\ref{sec_trigger} we calculate the impact parameter distribution 
in $pp$ collisions with production of a hard dijet near zero 
rapidity, and compare it with the impact parameter distribution 
for generic inelastic collisions. We demonstrate that 
hard dijet production acts as a ``centrality trigger''.
We also discuss the extension to production of multiple dijets
and the role of possible correlations in the transverse spatial
distribution of gluons. In Section~\ref{sec_bbl} we investigate the 
role of the ``centrality trigger'' in approaching the BBL
in central $pp$ collisions. We show that the trigger essentially 
eliminates collisions at large impact parameters where the soft 
interactions are not black. In Section~\ref{sec_final} we list the novel 
characteristics of the final state in central $pp$ collisions
which follow from the proximity to the BBL of the spectator 
parton interactions at small transverse distances.
Our conclusions are presented in Section~\ref{sec_conclusions}.
\section{Impact parameter distribution for generic inelastic
$pp$ collisions}
\label{sec_inelastic}
We begin by summarizing the available information about the
impact parameter distribution of the cross section for
generic inelastic $pp$ collisions at high energies. 
Most of our knowledge here comes from $pp$ elastic scattering, 
which has been studied in numerous experiments see Ref.~\cite{Block:1984ru} 
for a review. By unitarity (\textit{i.e.}, the optical theorem) the
$pp$ elastic amplitude contains information also about the total
(elastic plus inelastic) cross section, and thus about the
inelastic cross section.

It is well known from studies of $pp$ elastic scattering that 
the radius of strong interactions (the average impact parameter)
increases with the collision energy, $s$.  The $t$--slope of the 
elastic cross section, $B$, grows as
\be
B(s) \;\; =\;\; B(s_0) \; + \; 2 \alpha^{\prime}\; \ln (s/s_0),
\label{alpha_prime}
\ee
with $\alpha^{\prime} \approx 0.25 \, \text{GeV}^{-2}$. Thus, the 
radius of strong interactions is expected to be a factor of 1.5 
larger at LHC as compared to fixed target energies.

In the partonic picture, the mechanism for the increase of the 
radius of strong interactions with energy is the so-called 
Gribov diffusion. The emission processes in the 
soft parton ladder give rise to a random walk of partons in the 
transverse plane, reminiscent of a diffusion process 
\cite{Gribov:jg}. If one writes the amplitude of elastic 
scattering of two hadrons as a product of two $t$--dependent 
form factors, each parameterized by the transverse radius of the 
hadron, $R$, 
\be
A^{h_1 h_2} \;\; \propto \;\; \exp (t {R_1^2/4}) \; \exp (t {R_2^2/4}) ,
\ee
one can interpret the shrinkage of the diffractive cone as being 
due to an increase of the transverse spread of partons.
In terms of the average parton momentum fraction, $x$, 
this implies that the transverse area 
occupied by the low virtuality  partons in the hadron, $R^2$, 
increases with decreasing $x$ roughly as \cite{Gribov:jg}
\be
R^2(x) \;\; = \;\; R_0^2 \; + \; 2 \alpha^{\prime} \ln (x_0/x) . 
\ee

Detailed information about the distribution of the $pp$ cross section
(both elastic and inelastic) over impact parameters can be
obtained from the impact parameter representation of the
$pp$ elastic scattering amplitude, see e.g.\ Ref.~\cite{Block:1984ru}.
We write the invariant elastic amplitude in the form
\bea
A^{pp}(s, t) &=& \frac{i \, s}{4\pi} \int d^2 b \;
e^{-i (\bm{\Delta}_\perp \bm{b})}
\; \Gamma^{pp} (s, b)
\\
&=& \frac{i \, s}{2} \int_0^\infty db \, b \;
J_0 (\Delta_\perp b) \; \Gamma^{pp} (s, b) ,
\eea
where $\bm{\Delta}_\perp$ is a transverse momentum vector, with
$t = -\bm{\Delta}_\perp^2$ and $\Delta_\perp \equiv |\bm{\Delta}_\perp |$,
and $J_0$ denotes the Bessel function. Our normalization of the 
amplitude is the same as in Ref.~\cite{Islam:2002au}, \textit{cf.}\ the 
relation to the total cross section, Eq.~(\ref{sigma_tot_b}) below.
The dimensionless complex function $\Gamma^{pp}$ is called the profile 
function of the elastic amplitude. In this representation the integrated 
cross section for elastic scattering is given by 
\bea
\sigma^{pp}_{\text{el}} (s) &\equiv& \frac{4\pi}{s^2} \; \int^0_{-s} dt
\; |A^{pp}(s, t)|^2
\\
&=& \int d^2 b \; |\Gamma^{pp} (s, b)|^2 .
\label{sigma_el_b}
\eea
In the last step, we have used that at large $s$ the lower limit
of the $t$--integral can be replaced by $-\infty$, and that 
for an amplitude independent of the azimuthal scattering angle
\be
\int_{-\infty}^0 dt \;\; = \;\; 4\pi \; 
\int\frac{d^2 \Delta_\perp}{(2\pi )^2} .
\ee
A similar representation can be derived for the total cross section for 
$pp$ scattering, which by the optical theorem is
proportional to the imaginary part of the forward ($t = 0$) elastic 
amplitude:
\bea
\sigma^{pp}_{\text{tot}} (s) &=& \frac{8\pi}{s} \; \text{Im} 
\; A^{pp}(s, t = 0)
\\
&=&
2 \int d^2 b \; \text{Re}\; \Gamma^{pp} (s, b) .
\label{sigma_tot_b}
\eea
Finally, taking the difference of Eqs.~(\ref{sigma_tot_b})
and (\ref{sigma_el_b}), one obtains a representation of the
inelastic (total minus elastic) $pp$ cross section as an
integral over impact parameters:
\bea
\sigma^{pp}_{\text{in}} (s) &\equiv& \sigma^{pp}_{\text{tot}} (s) 
- \sigma^{pp}_{\text{el}} (s)
\nonumber \\
&=& \int d^2 b \; \left[ 
2 \text{Re}\; \Gamma^{pp} (s, b) - |\Gamma^{pp} (s, b)|^2 \right] .
\label{sigma_in_b}
\eea

\begin{figure}
\includegraphics[width=8cm,height=8cm]{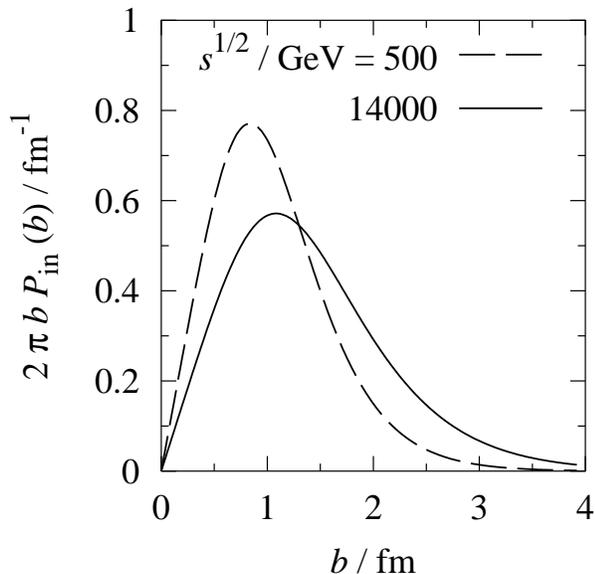}
\caption[]{The normalized impact parameter distribution for generic 
inelastic collisions, $P_{\text{in}} (s, b)$, Eq.~(\ref{P_in_def}),
obtained with the parameterization of the elastic $pp$ amplitude
of Islam \textit{et al.} \cite{Islam:2002au}
(``diffractive'' part only). The plot shows the 
``radial'' distribution in the impact parameter
plane, $2 \pi b \, P_{\text{in}} (s, b)$. The energies are 
$\sqrt{s} = 500\, \text{GeV}$ (RHIC) and
$14000\, \text{GeV}$ (LHC).}
\label{fig_P_in}
\end{figure}
The integrand of Eq.~(\ref{sigma_in_b}) represents the distribution 
of the cross section for generic inelastic collisions (\textit{i.e.}, 
summed over all inelastic final states) over impact parameters.
It is convenient to define a normalized $b$--distribution as
\be
P_{\text{in}} (s, b) \;\; = \;\; 
\frac{2 \text{Re}\; \Gamma^{pp} (s, b) - |\Gamma^{pp} (s, b)|^2}
{\sigma_{\text{in}} (s)} .
\label{P_in_def}
\ee
For a quantitative estimate of this distribution we can use
phenomenological parameterizations of the $pp$ elastic scattering 
amplitude, which fit the presently available $pp$ elastic
data at collider energies. The results obtained with the
parameterization of of Islam \textit{et al.} \cite{Islam:2002au}
(``diffractive'' part only) are presented in Fig.~\ref{fig_P_in},
for energies corresponding to RHIC and the LHC.
It should be noted that the predictions for LHC are based 
on extrapolation of fits to the presently available data over 
nearly two orders of magnitude in $s$. The biggest uncertainty in
the extrapolation appears to be due to the uncertainties in the
measurement of $\sigma^{\bar p p}_{\text{tot}}$ at the Tevatron and
the limited range of $t$ covered in the collider measurements of
elastic $\bar p p$ scattering.

In principle one should include here also effects of inelastic diffraction. 
However, this contribution should be rather small at LHC energies 
due to the blackening of interaction, see Ref.~\cite{Kaidalov:2003vg} for a 
recent review. Besides this, one expects that a significant part of 
inelastic diffraction at $t < 0$ is due to the spin flip 
amplitudes \cite{Alberi:af}.
\section{Transverse spatial distribution of hard partons in the nucleon}
\label{sec_hard}
We shall now review what is known about the transverse 
distribution of hard partons in the nucleon. Based on this, we 
shall later proceed to estimate the impact parameter distribution
in $pp$ collisions with hard dijet production.

Numerous measurements of hard inclusive scattering processes
(DIS, Drell--Yan pair production) have produced a rather detailed 
picture of the longitudinal momentum distribution of partons in the 
nucleon. The study of the transverse spatial distribution of partons 
is still at a much more primitive stage. Information about the
transverse spatial distribution of gluons is contained in
the so-called two--gluon form factor of the nucleon, which 
parametrizes the $t$--dependence of the (generalized) gluon
distribution in the nucleon,
\be
g(x, t; Q^2) \;\; = \;\; g(x; Q^2) \; F_g (x, t; Q^2) ,
\label{twogluon_def}
\ee
where
\be
F_g (x, t = 0; Q^2) \;\; = \;\; 1 ,
\ee
and $g(x; Q^2)$ is the usual gluon distribution in the nucleon.
We define the Fourier transform of this form factor as
\be
F_g (x, \rho; Q^2) \; \equiv \; \int \frac{d^2 \Delta_\perp}{(2 \pi)^2}
\; e^{i (\bm{\Delta}_\perp \bm{\rho})}
\; F_g (x, t = -\bm{\Delta}_\perp^2; Q^2) ,
\label{rhoprof_def}
\ee
where $\bm{\rho}$ is a two--dimensional coordinate variable,
and $\rho \equiv |\bm{\rho}|$ (this variable is named $\bm{b}$
in Ref.~\cite{Strikman:2003gz}; in this paper we reserve $\bm{b}$ 
for the impact parameter vector of the $pp$ collision). This function 
describes the spatial distribution of gluons in the transverse plane. 
It is normalized to unit integral over the transverse plane,
\be
\int d^2 \rho \; F_g (x, \rho; Q^2 ) \;\; = \;\; 1.
\ee
A measure of the transverse size of the nucleon for given $x$ and $Q^2$
is the average of $\rho^2$ calculated with this distribution,
which is identical to 4 times the $t$--slope of the two--gluon 
form factor at $t = 0$,
\bea
\langle \rho^2 \rangle (x, Q^2) &\equiv&
\int d^2 \rho \; \rho^2 \; F_g (x, \rho; Q^2 )
\\
&=&
4 \, \left. \frac{\partial}{\partial t} \, F_g (x, t; Q^2) \; 
\right|_{t = 0} .
\eea

For sufficiently small $x \; (\alt 0.3)$ the parameter $\rho$ can 
be interpreted as the distance of the parton from the 
center of mass of the nucleon in the transverse plane. 
For larger $x$ the interpretation of 
the $\rho$--distribution becomes less intuitive, as in this case 
one can no longer neglect the difference between the longitudinal 
momentum of the spectator system and that of the whole nucleon.
In the limit $x \to 1$ the active parton would carry the 
entire longitudinal momentum of the nucleon, and only soft 
partons would be left in the spectator system, see 
Ref.~\cite{Burkardt:2002hr} for a discussion.
This shall not concern us here, since we shall be interested
in the gluon distribution at $x \leq 0.05$, as relevant for
hard dijet production in the central rapidity region (see below).

%
%
\begin{figure}
\includegraphics[width=8cm,height=8cm]{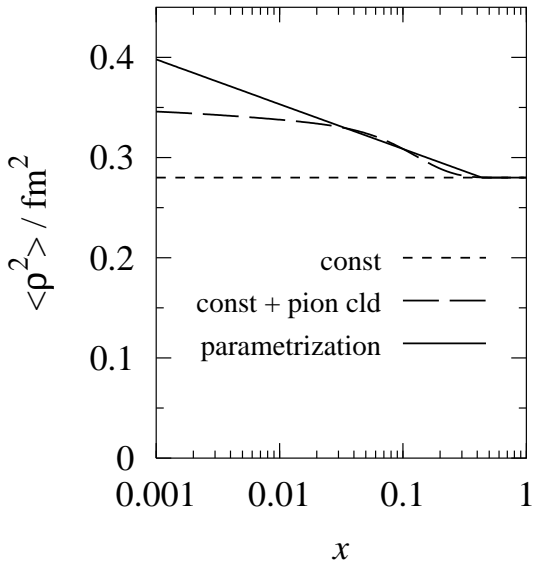}
\caption[]{Our model for the $x$--dependence of the average 
transverse gluonic size squared of the nucleon, $\langle \rho^2 \rangle$
at the scale $Q_0^2 = 2 \div 4 \, \text{GeV}^2$ 
relevant to $J/\psi$ production.
\textit{Short--dashed line:} $\langle \rho^2 \rangle = 0.28 \, \text{fm}^2$,
as extracted from the $t$-slope of the $J/\psi$ production
cross section measured in various experiments \cite{Frankfurt:2002ka}.
\textit{Long--dashed line:} Sum of the constant value 
$\langle \rho^2 \rangle = 0.28 \, \text{fm}^2$
and the pion cloud contribution calculated in Ref.~\cite{Strikman:2003gz}.
\textit{Solid line:} The parameterization Eq.~(\ref{rho2_x_simple}),
based on the experimental value of 
$\alpha^{\prime}_{\text{hard}}$, 
Eq.~(\ref{alphap_hard}) \cite{Chekanov:2002xi}.}
\label{fig_param}
\end{figure}
At moderately small $x \; (\agt 0.001 )$, it is possible to obtain 
information about the two--gluon form factor at a resolution scale of
$Q_0^2 \sim 2 \div 4 \, \text{GeV}^2$ from the analysis of exclusive
$J/\psi$ photo (or electro--) production off the 
nucleon \cite{Frankfurt:2002ka}.
It turns out that for $x\geq 0.01$ this form factor is significantly 
harder than the electromagnetic form factor of the nucleon. 
The $t$--dependence of the cross section is well described 
by a dipole form factor \cite{Frankfurt:2002ka}
\be
F_g (x, t) \;\; = \;\; \frac{1}{( 1 - t/m_{g}^2)^2} ,
\label{dipole}
\ee
with a mass parameter 
\be
m_{g}^2 \;\; \approx \;\; 1.1\, \text{GeV}^2 \;\; \gg \;\; m_\rho^2
\hspace{3em} (x \agt 0.01).
\label{m_g_const}
\ee
The corresponding spatial distribution of gluons in the transverse plane, 
Eq.~(\ref{rhoprof_def}), is given by
\bea
F_g (x, \rho ) &=& \frac{m_{g}^2}{2\pi}
\; \left(\frac{m_{g} \rho}{2}\right) \; K_1 (m_{g} \rho ) ,
\label{dipole_b}
\eea
where $K_1$ denotes the modified Bessel function. Note that 
this function is positive, in agreement with the general positivity
condition for the transverse coordinate--dependent gluon density,
derived in Ref.\cite{Pobylitsa:2002iu}. The average 
$\langle \rho^2 \rangle$ corresponding to this distribution is inversely 
proportional to the mass parameter squared,
\bea
\langle \rho^2 \rangle &=& \frac{8}{m_{g}^2} .
\label{rho2_from_m2}
\eea
Numerically, the value (\ref{m_g_const}) amounts to 
\be
\langle \rho^2 \rangle \;\; \approx \;\; 0.28 \, \text{fm}^2
\hspace{3em} (x \agt 0.01),
\ee
which is a factor of $\sim 1.5$ smaller than the proton's ``transverse 
electric charge radius squared'', 
$(2/3) \langle r^2 \rangle_{\text{em}}$.

For smaller values of $x \; (0.01 \alt x \alt 0.1)$ the
transverse gluonic size of the nucleon starts to increase.
This can be explained semi-quantitatively by the kicking in 
of contributions from the pion cloud of the nucleon, which 
are suppressed for $x > M_\pi / M_N$ \cite{Strikman:2003gz}, 
see Fig.~\ref{fig_param}. The 
analysis of $J/\psi$ photoproduction data has shown that 
the size keeps growing also for $x \alt 0.01$ \cite{Chekanov:2002xi}. 
In all, the rate of the increase of the gluonic size between 
$x\sim 10^{-2}$ and $10^{-3}$ is about a factor of two 
\textit{smaller} than that of the total cross section, which 
is dominated by soft physics (see Section~\ref{sec_inelastic}):
\bea
\frac{1}{4} \; \frac{\partial \; \langle \rho^2 \rangle}{\partial\ln (1/x)}
&\equiv& \alpha^{\prime}_{\text{hard}} \;\; \approx \;\;
0.125 \, \text{GeV}^{-2}
\nonumber \\
&& (x = 10^{-3} \div 10^{-2}),
\label{alphap_hard}
\eea
which should be compared with Eq.~(\ref{alpha_prime}).
We can pa\-ra\-me\-trize the $x$--dependence of $\langle \rho^2 \rangle$, 
at the scale $Q_0^2$ probed in $J/\psi$ production, by combining this 
experimentally determined rate of increase
with our model estimate of $\langle \rho^2 \rangle$
(constant plus pion cloud contribution) at $x = 0.1$,
$\langle \rho^2 \rangle = 0.31 \, \text{fm}^2$ \cite{Strikman:2003gz}.
This amounts to the parametrization
\be
\langle \rho^2 \rangle (x, Q_0^2) \;\; = \;\; \text{max} \;
\left\{\begin{array}{ll}
0.31 \, \text{fm}^2 \; + \;  0.0194 \, \text{fm}^2 \; 
\ln \frac{\displaystyle 0.1}{\displaystyle x}
\\[2ex]
0.28 \, \text{fm}^2
\end{array} \right. .
\label{rho2_x_simple}
\ee
This simple form fits well the 
$x$--dependence of $\langle \rho^2 \rangle$ due to pion cloud 
contributions in the region $0.01 \leq  x \leq 0.1$, 
as calculated in Ref.~\cite{Strikman:2003gz}, see Fig.~\ref{fig_param}, 
and continues it down to smaller values of $x$ using the experimentally 
measured rate of increase. 

For our estimates of the probability of hard multijet production
we need to model not only the average transverse size of the nucleon,
but the full transverse spatial distribution of gluons. 
For simplicity, we shall assume that at the scale 
$Q_0^2$ the $\rho$--distribution
at all relevant $x$ can be described by the Fourier transform 
of a dipole form factor, \textit{cf.}\ Eq.~(\ref{dipole_b}), but with an 
$x$--dependent mass parameter. This parameter is then 
uniquely determined by the value of $\langle\rho^2\rangle$, as
given by Eq.~(\ref{rho2_x_simple}), via Eq.~(\ref{rho2_from_m2}).
This defines our model of the $x$-- and $\rho$--dependent
distribution of gluons at the scale $Q_0^2$.
 
We are interested in the $\rho$--dependent gluon distribution 
at large virtualities, corresponding to hard dijet production 
at LHC. This requires to take into account the effect of 
DGLAP evolution on the $\rho$--dependent distributions. The 
$Q^2$--evolution of the parton distributions is diagonal in 
$\rho$. It degrades the longitudinal momentum fractions $x$ of 
the partons, resolving a parton with given $x$ into a collection 
of partons with smaller $x$, while the transverse location of the 
new partons is practically the same as that of the ``parent''
parton provided $Q\gg 1/\rho$. Nevertheless, $Q^2$ evolution 
does change the $\rho$--profile of the distributions at given 
$x$. Generally speaking, evolution will reduce the rate of 
broadening of the distributions with decreasing of $x$. This 
happens because with increasing $Q^2$ the parton distributions 
at the high scale become sensitive to the input distribution at the low 
scale $Q_0^2$ at larger and larger $x$--values, where their transverse 
size becomes small. 

%
%
\begin{figure}
\includegraphics[width=8cm,height=8cm]{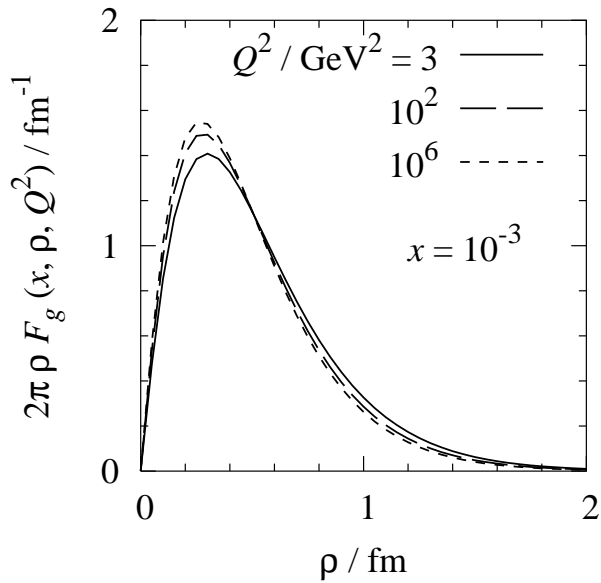}
\caption[]{The change of the normalized $\rho$--profile of the gluon 
distribution, $F_g (x, \rho; Q^2)$, Eq.~(\ref{rhoprof_def}), 
with $Q^2$, as due to DGLAP evolution, for $x = 10^{-3}$. 
The input gluon distribution is the Gl\"uck--Reya--Vogt 
parametrization at $Q_0^2 = 3 \, \text{GeV}^2$, with a dipole--type
$\rho$--profile, Eq.~(\ref{dipole_b}), of size determined by 
the parametrization Eq.~(\ref{rho2_x_simple}).}
\label{fig_evol_profile}
\end{figure}
%
%
\begin{figure}
\includegraphics[width=8cm,height=8cm]{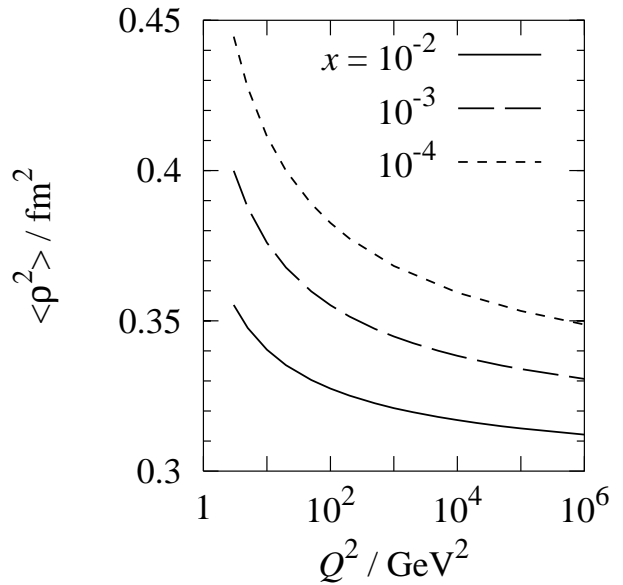}
\caption[]{The change of the average transverse gluonic size 
squared, $\langle \rho^2 \rangle$, due to DGLAP evolution,
for $x = 10^{-2}, 10^{-3}$ and $10^{-4}$.}
\label{fig_evol_rho2}
\end{figure}
We have studied numerically the leading--order $Q^2$--evolution 
of our model of the $\rho$--dependent gluon distribution at the scale 
$Q_0^2$ \textit{cf.}\ Eqs.~(\ref{dipole_b}), 
(\ref{rho2_x_simple}) and (\ref{rho2_from_m2}), employing the numerical 
method described in Ref.~\cite{Kwiecinski:gc}.
We have used the Gl\"uck--Reya--Vogt leading--order 
parameterization \cite{Gluck:1998xa} to describe 
the total (integrated over $\rho$) gluon and singlet quark distributions
at the input scale $Q_0^2$ and modeled their $\rho$--profile as described 
above. The value of the input scale we have taken as 
$Q_0^2 = 3 \, \text{GeV}^2$, which is the central value of the range 
of scales associated with $J/\psi$ photoproduction. For simplicity 
we assume identical $\rho$--profiles for the gluon and 
singlet quark distributions at the input scale. 
The results of the DGLAP evolution of the $\rho$--dependent gluon
distribution are shown in Figs.~\ref{fig_evol_profile}
and Fig.~\ref{fig_evol_rho2}.
Fig.~\ref{fig_evol_profile} shows the normalized $\rho$--profile,
$F_g (x, \rho; Q^2)$, \textit{cf.} Eq.~(\ref{rhoprof_def}),
corresponding to the distribution after evolving to the higher 
scale $Q^2$, for the value $x = 10^{-3}$.
Fig.~\ref{fig_evol_rho2} shows the $Q^2$--dependence of the
average size squared, $\langle \rho^2 \rangle$, as induced by DGLAP 
evolution, for values of $x = 10^{-2}, 10^{-3}$ and $10^{-4}$. 
The corresponding curves for the
singlet quark distribution would be qualitatively similar.
One can see that the effect of transverse 
broadening of the gluon distribution with decreasing $x$, 
which is rather small already at the initial resolution scale, 
is further reduced at the scale relevant for the LHC kinematics 
($Q \geq 20 \, \text{GeV}$). 

Fig.~\ref{fig_evol_profile} demonstrates that the deviations
of the $\rho$--profile from the dipole shape at the initial scale, 
Eqs.~(\ref{dipole_b}), due to DGLAP evolution are very small. This suggests 
a simplified parametrization of the combined $Q^2$-- and $x$-- 
dependence of the $\rho$--profile of the gluon distribution, 
in which the dipole shape is assumed to hold at all $Q^2$ (and $x$),
and the $Q^2$-- (and $x$--) dependence is entirely ascribed to the
mass parameter $m_{g}^2$. We fit the combined $Q^2$-- and $x$--dependence 
of $\langle\rho^2\rangle$ due to DGLAP evolution in the region 
$Q_0^2 \le Q^2 \le  10^6 \, \text{GeV}^2$, as shown in 
Fig.~\ref{fig_evol_rho2}, by the simple two--parameter form
\be
\langle\rho^2\rangle (x, Q^2)
\;\; = \;\; \langle\rho^2\rangle (x, Q_0^2)
\; \left( \displaystyle 1 + A\; \ln\frac{Q^2}{Q_0^2} 
\right)^{\textstyle -a} ,
\label{rho2_qdep}
\ee
where $Q_0^2 = 3 \, \text{GeV}^2$, 
$\langle\rho^2\rangle (x, Q_0^2)$ is defined by 
Eq.~(\ref{rho2_x_simple}), and 
\be
A \;\; = \;\; 1.5 , 
\hspace{3em}
a \;\; = \;\; 0.0090 \; \ln \frac{1}{x} .
\ee
For each $x$ and $Q^2$, this value of $\langle\rho^2\rangle$ defines
a dipole mass parameter $m_{g}^2$ via Eq.~(\ref{rho2_from_m2}).
Our model for the $\rho$--dependent gluon distribution is then
given by the Gl\"uck--Reya--Vogt leading--order parametrization 
for the total distribution at the scale $Q^2$, times the normalized 
dipole $\rho$--profile, Eq.~(\ref{dipole_b}), 
with this mass parameter. This parametrization has the 
correct $Q^2$--dependence of the transverse size is ``built in'',
removing the need to perform explicit DGLAP evolution of the 
$\rho$--dependent distributions, which is very convenient for
the following studies.

\section{Impact parameter distribution for a hard multijet trigger}
\label{sec_trigger}
Using the information about the transverse distribution
of partons in the proton, we can now investigate the impact parameter 
dependence of the cross section for inelastic collisions with production 
of two jets in the central rapidity region. In particular,
we shall show that a trigger on hard dijet events allows to
reduce the effective impact parameter of $pp$ collisions
as compared to generic inelastic collisions studied
in Section~\ref{sec_inelastic}. 

\begin{figure}
\includegraphics[width=4cm,height=4cm]{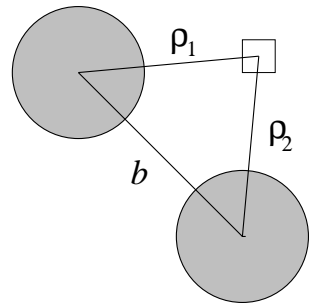}
\caption[]{Illustration of the overlap integral of parton
distributions in the transverse plane, defining the $b$--distribution 
for binary parton collisions producing a dijet, Eq.~(\ref{P_2_def}).}
\label{fig_overlap}
\end{figure}

We consider the production of two jets (with equal but opposite
transverse momentum) in a binary parton--parton
collision. The resolution scale is given by the transverse momentum
squared of one the jets, $q_\perp^2$. The momentum fractions of the two 
colliding partons with respect to their parent protons, $x_1$ and $x_2$, 
can be reconstructed from the measured energy and momenta of the two jets. 
Four--momentum conservation implies for the scattering at $90^\circ$ at the 
center of mass of two partons
\be
x_1 \, x_2 \;\; = \;\; \frac{4 q_\perp^2}{s} ,
\label{x1_x2}
\ee
where $4 q_\perp^2$ is the invariant mass squared of the two--jet system.
In the following we shall be interested in two jets near zero 
total rapidity, which requires 
\be
x_1 \;\; \approx \;\; x_2 .
\label{x_dijet}
\ee 
Under this condition the momentum fractions are completely fixed
by Eq.~(\ref{x1_x2}). In the following we consider the dijet production 
due to collision of two gluons, since such partonic collisions give 
the dominant contribution to the total cross section.
The probability for a binary collision of two gluons is proportional
to the product of the gluon densities in transverse space in the two 
colliding protons, taking into account that their transverse centers
are separated by a distance $b$ --- the impact parameter of the
$pp$ collision, see Fig.~\ref{fig_overlap}. This implies that
the distribution of the cross section for such events over
the impact parameter $b$ is given by
\bea
P_2 (b) &\equiv&
\int d^2\rho_1 \int d^2\rho_2 \; 
\delta^{(2)} (\bm{b} - \bm{\rho}_1 + \bm{\rho}_2 )
\nonumber \\
&\times& F_g (x_1, \rho_1 ) \; F_g (x_1, \rho_2 ) ,
\label{P_2_def}
\eea
where $x_1 = 2 q_\perp / \sqrt{s}$, \textit{cf.}\ Eqs.~(\ref{x1_x2}) and 
(\ref{x_dijet}), and the scale of the gluon $\rho$--profiles is $q_\perp^2$.
This distribution is normalized such that the integral over all 
$\bm{b}$ is unity. Since it has the form of a convolution in 
the parton transverse positions, it can also be expressed as the Fourier 
transform of the square of the two--gluon form factor, Eq.~(\ref{dipole}).
In particular, for the two--gluon form factor of dipole form, 
Eq.~(\ref{dipole_b}), used in our model of the $\rho$--dependent
gluon distribution (see Section~\ref{sec_hard}) one obtains
\be
P_2 (b) \;\; = \;\; \frac{m_{g}^2}{12\pi} 
\left( \frac{m_{g} b}{2} \right)^3  K_3 (m_{g} b) ,
\label{P_2_dipole}
\ee 
where $m_{g}$ should be substituted by the value corresponding to
$x_1 = 2 q_\perp / \sqrt{s}$ and $Q^2 = q_\perp^2$, 
see Eq.~(\ref{rho2_qdep}).
%
%
%

%
%
\begin{figure}[t]
\includegraphics[width=8cm,height=8cm]{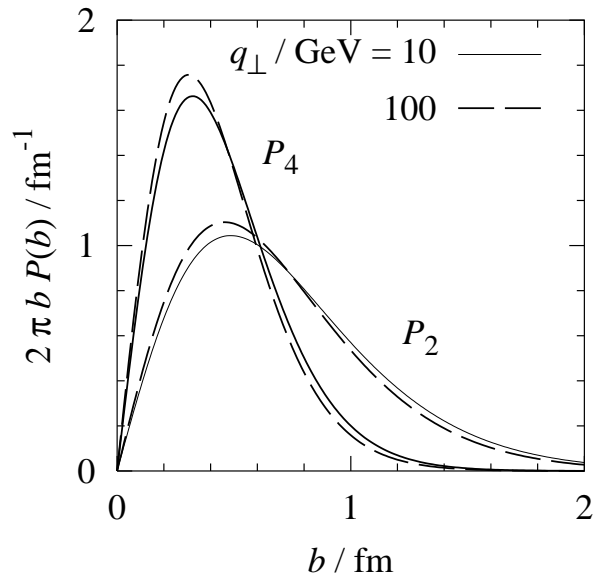}
\caption[]{The $b$--distribution for the trigger on 
hard dijet production, $P_2 (b)$, obtained with the dipole
form of the gluon $b$--profile, Eq.~(\ref{P_2_dipole}),
for $\sqrt{s} = 14000\, \text{GeV}$ and
$q_\perp = 10\, \text{GeV}$ and $100\, \text{GeV}$.
The plots show the ``radial'' distributions in the impact 
parameter plane, $2 \pi b \, P_2 (b)$.
Also shown is the corresponding distribution for a
trigger on double dijet production, $P_4(b)$, with the same $p_\perp$.}
\label{fig_P2_pt}
\end{figure}
Fig.~\ref{fig_P2_pt} shows the distribution $P_2 (b)$ 
for a center--of--mass energy of $\sqrt{s} = 14000\, \text{GeV}$ (LHC), 
and two values of the jet momentum, 
$q_\perp = 10\, \text{GeV}$ and $100\, \text{GeV}$.
One sees that the distribution is rather insensitive to the 
precise value of the jet momentum. 
This can be explained by the relatively slow decrease of 
$\langle \rho^2\rangle $ with increasing $x$ and $Q^2$.
The average values of impact
parameter squared, $\langle b^2 \rangle$, calculated with these
distributions, is $0.71 \, \text{fm}^2$ for $q_\perp = 10\, \text{GeV}$ 
and $0.63 \, \text{fm}^2$ for $q_\perp = 100\, \text{GeV}$.

In Fig.~\ref{fig_P2_pt} we assume production of a two--jet system
at zero rapidity, \textit{cf.}\ Eq.~(\ref{x_dijet}). If we considered
instead a two--jet system at some non-zero rapidity, $y$, the
(anyway weak) dependence of the $\rho$--distributions in
Eq.~(\ref{P_2_def}) on $x_1$ and $x_2$ would work in opposite 
directions, leading to an extremely weak dependence of our results 
on the rapidity of the produced system over a wide range of $y$.

\begin{figure*}
\begin{tabular}{cc}
\includegraphics[width=8cm,height=8cm]{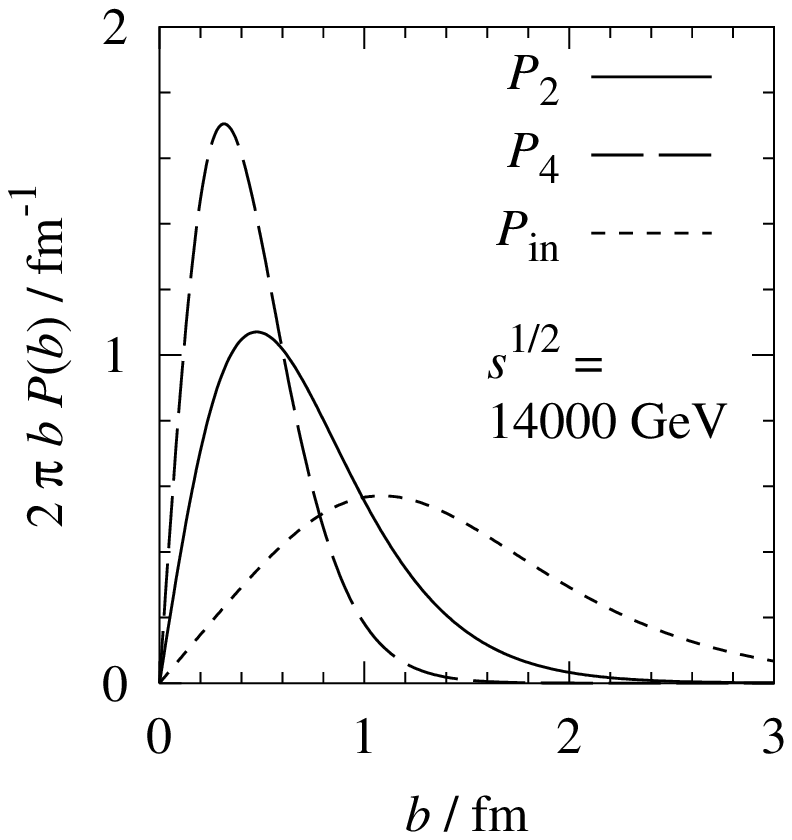}
&
\includegraphics[width=8cm,height=8cm]{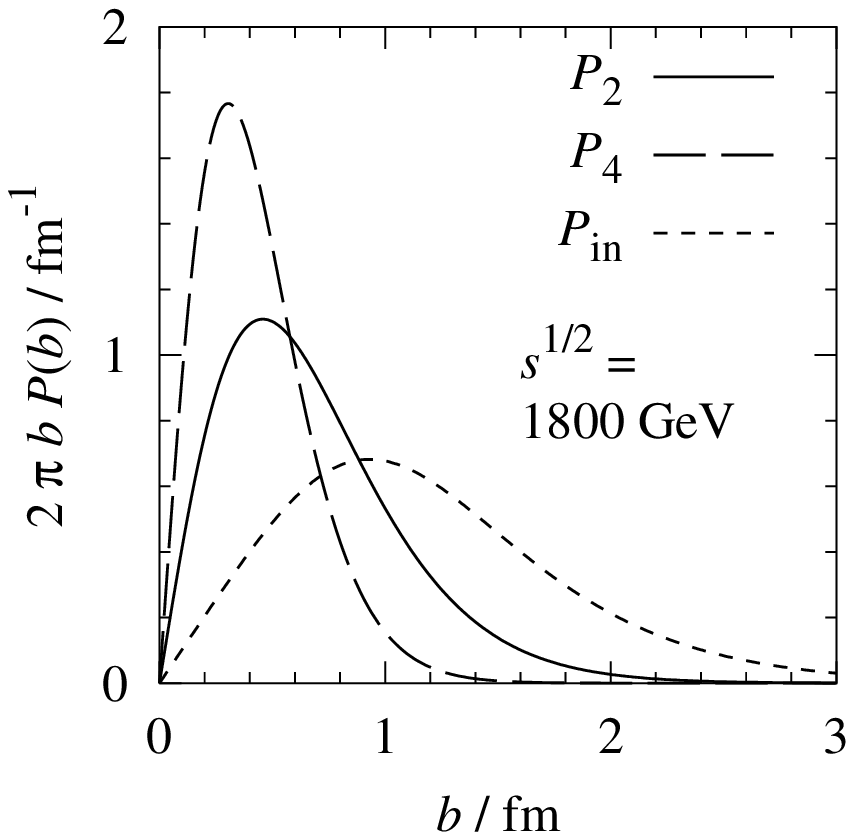}
\\
\includegraphics[width=8cm,height=8cm]{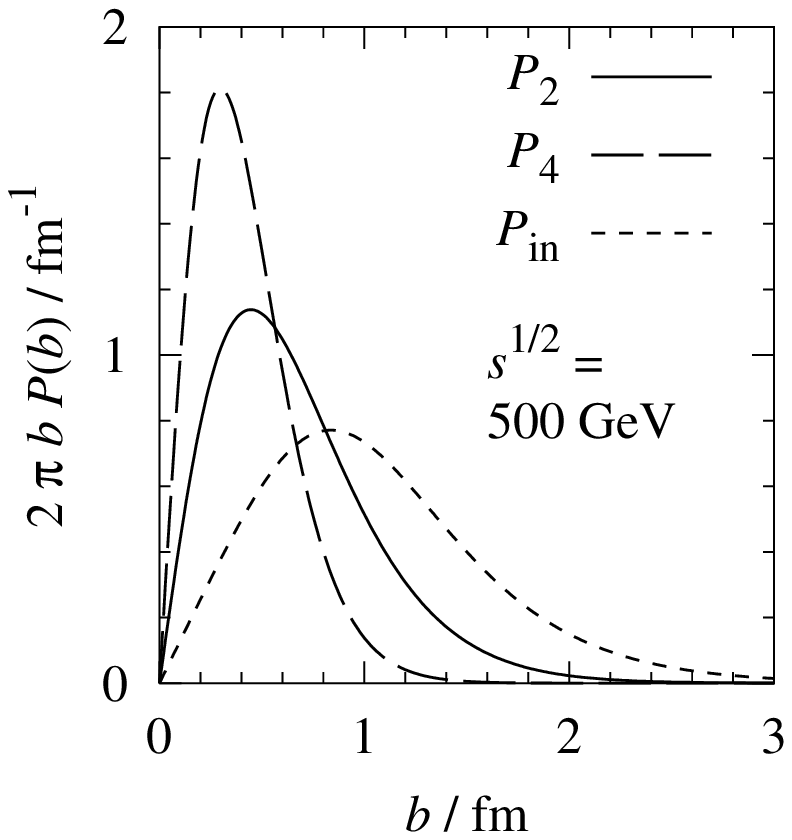}
&
\end{tabular}
\caption[]{\textit{Solid lines:} $b$--distributions for the 
dijet trigger, $P_2 (b)$, with $q_\perp = 25 \, \text{GeV}$,
as obtained from the dipole--type gluon $\rho$--profile,
Eq.~(\ref{P_2_dipole}). \textit{Long--dashed line:} 
$b$--distribution for double dijet events, $P_4 (b)$. 
\textit{Short--dashed line}: $b$--distribution 
for generic inelastic collisions, obtained from the parameterization 
of the elastic $pp$ amplitude of Islam \textit{et al.} \cite{Islam:2002au}
(``diffractive'' part only), \textit{cf.}\ Fig.~\ref{fig_P_in}. 
Shown are the results for $\sqrt{s} = 14000 \, \text{GeV}$ (LHC),
$1800 \, \text{GeV}$ (Tevatron $\bar p p$), and $500 \, \text{GeV}$ (RHIC). 
The plots show the ``radial'' distributions in the impact parameter
plane, $2 \pi b \, P (b)$.} 
\label{fig_pb}
\end{figure*}

In Fig.~\ref{fig_pb} we compare the $b$--distribution for the
hard dijet trigger, $P_2 (b)$ (solid line), 
with the $b$--distribution for generic inelastic
events, $P_{\text{in}} (s)$, estimated in Section~\ref{sec_inelastic}. 
The short--dashed line in Fig.~\ref{fig_pb} shows the distribution 
$P_{\text{in}} (s)$ obtained from the parameterization of the elastic 
$pp$ amplitude of Islam \textit{et al.} \cite{Islam:2002au} (``diffractive'' 
part only). Shown are the results for $\sqrt{s} = 14000 \, \text{GeV}$ (LHC),
$1800 \, \text{GeV}$ (Tevatron $\bar p p$), and $500 \, \text{GeV}$ (RHIC). 
A momentum of $q_\perp = 25 \, \text{GeV}$ was assumed for the dijet trigger.
One sees that in all cases the $b$--distribution for dijet events 
is much narrower than the one for generic inelastic collisions. 
The corresponding averages $\langle b^2 \rangle$ are given in 
Table~\ref{table_pb}. The average $\langle b^2 \rangle$
for the hard dijet trigger rises much more slowly with 
$s$ than for generic inelastic collisions, which are dominated
by soft physics. Thus, the reduction in effective impact 
parameters due to the dijet trigger is most pronounced at
LHC energies, where $\langle b^2 \rangle$ is reduced to
$\sim 1/4$ its value for generic inelastic collisions.
%
%

A further reduction of the effective impact parameters can be achieved
by a trigger on events with two dijets, \textit{i.e.}, two binary hard
parton collisions (such processes can be easily distinguished from 
the leading twist $2 \to 4$ processes in the collider experiments, 
see \textit{e.g.} Ref.~\cite{Abe:1997bp}). 
It was estimated in Ref.~\cite{Lippmaa:1997qb} that
this reduces $\left<b^2\right>$ by a factor of two as compared to 
the single dijet trigger. In our approach, the $b$--distribution for the
double dijet trigger is given by 
\be
P_4 (b) \;\; = \;\; \frac{P_2^2(b)}{\displaystyle 
\int d^2 b \; P_2^2(b)}.
\ee
For simplicity we assume here identical $x_1$ and $q_\perp$ for the two
dijets; the definition could easily be generalized to allow for
different values. For the two--gluon form factor of dipole form, 
Eq.~(\ref{dipole_b}), this becomes
\be
P_4 (b) \;\; = \;\; \frac{7 \, m_{g}^2}{36\pi} 
\left( \frac{m_{g} b}{2} \right)^6  \left[ K_3 (m_{g} b) \right]^2 ,
\label{P_4_dipole}
\ee 
where again $m_{g}$ should be substituted by the value corresponding to
$x_1 = 2 q_\perp / \sqrt{s}$ and $Q^2 = q_\perp^2$, 
see Eq.~(\ref{rho2_qdep}).
Fig.~\ref{fig_P2_pt} shows that the $b$--distribution for the 
double dijet trigger, $P_4 (b)$ is equally insensitive to the precise
value of $q_\perp$ as that for the dijet trigger, $P_2 (b)$.
The comparison in Fig.~\ref{fig_pb} and Table~\ref{table_pb}
shows that the double dijet trigger allows for a further reduction
of the effective impact parameters by a factor of
$\sim$ 2.5 compared to the dijet trigger.

In calculating $P_4(b)$ we have made the assumption that the gluons 
are not correlated in the transverse plane. To test this assumption we
can  compare the rate of double dijet production in our model to the one
which is observed in the CDF experiment \cite{Abe:1997bp}. 
The ratio of the cross section of double dijet events and the square of 
the single dijet cross section is proportional 
to \cite{Abe:1997bp,Calucci:1999yz}
\be
\sigma_{\text{eff}} \;\; = \;\; \left[ \int d^2 b \; P_2^2(b) \right]^{-1} .
\ee
In our calculation  we find $\sigma_{\text{eff}} = 34 \, \text{mb}$, 
which should be compared with
$\sigma_{\text{eff}} = 14.5\pm1.7^{+1.7}_{-2.3} \, \text{mb}$ 
reported by CDF assuming that there is no 
correlations in the longitudinal distribution of partons. 
(In Ref.~\cite{Calucci:1999yz} $\sigma_{\text{eff}} \sim 30\, \text{mb}$ 
was obtained assuming that the parton distribution similar to that of 
valence quarks, a hypothesis resembling the conclusion we derived from 
the leading twist analysis of the $J/ \psi$ elastic 
photoproduction \cite{Frankfurt:2002ka}.) A factor of two difference 
between the theoretical number and the data may indicate that 
there are significant  transverse correlations in the parton density 
at the resolution  scale of $Q \geq 5\, \text{GeV}$ probed by CDF.
Such correlations could result, for example, due to the DGLAP 
evolution from a low $Q^2$ scale of a couple $\text{GeV}^2$ to 
$Q \sim 25\, \text{GeV}^2$, since the partons 
emitted in the course of this evolution would have small transverse
separation (the so--called ``hot spots'' of Ref.~\cite{Mueller:ey}).
Assuming that the difference between the uncorrelated model and 
the data is due to such \textit{local correlations in $b$} we would
obtain a $b$-distribution for the double dijet trigger 
approximately as
\bea 
P_{4, \text{corr}}(b) &\approx&
P_2(b) \; 
\frac{\sigma_{\text{eff}} (\text{model}) - \sigma_{\text{eff}} (\text{CDF})}
{\sigma_{\text{eff}} (\text{model})}
\nonumber \\
&+& P_4(b) \;
\frac{\sigma_{\text{eff}}(\text{CDF})}{\sigma_{\text{eff}} (\text{model})} ,
\eea
where $P_2(b)$ and $P_4(b)$ are the above uncorrelated model estimates.
It is clear from the inspection of Fig.~\ref{fig_pb} that this still 
corresponds to a large reduction (by a factor $\sim 1.5$)
of the effective impact parameters with the double dijet trigger 
as compared to the single dijet trigger.
%
%
\begin{table}
\[
\begin{array}{|c|c|c|c|c|}
\hline
\text{Facility} & \sqrt{s} / \text{GeV} & 
\langle b^2 \rangle_2 / \text{fm}^2 & \langle b^2 \rangle_4 / \text{fm}^2 & 
\langle b^2 \rangle_{\text{in}} / \text{fm}^2 
\\ \hline
\text{LHC}      & 14000  & 0.67 & 0.26 & 2.7  \\
\text{Tevatron} & 1800 & 0.63 & 0.24 & 1.8  \\
\text{RHIC}     & 500 & 0.59 & 0.23 & 1.43 \\
\hline
\end{array}
\]
\caption[]{The average impact parameter squared, $\langle b^2 \rangle$,
corresponding to the $b$--distributions $P_2 (b), P_4(b)$, and 
$P_{\text{in}}(b)$, shown in Fig.~\ref{fig_pb}.}
\label{table_pb}
\end{table}

It should be noted that the jet cross sections discussed here 
are sensitive to the choice of ``primordial'' transverse momentum
distribution of the colliding partons, to higher--order radiative corrections, 
and to the jet definition adopted. To our knowledge, the same procedure 
was used in analysing single and double dijet events in 
Ref.~\cite{Abe:1997bp} (one of the jets was taken to be a photon
for simplicity), so most of these effects should cancel in the 
cross section ratio. An additional source of uncertainty is the
correlation of the soft background in hadron production with the
centrality of the $pp$ event, as discussed in Section~\ref{sec_final}.
A quantitative study of parton--parton correlations certainly requires 
a more careful investigation of these effects.

To summarize, we have demonstrated that a trigger on events with
(one or more) dijets near zero rapidity strongly reduces the effective 
impact parameters in $pp$ collisions at LHC energies. 
Such a trigger can thus be used as a ``centrality filter''. 
This is of considerable practical interest, as the characteristics of 
the final state strongly depend on the centrality of the $pp$ collision.
\section{Approaching the black--body limit in central pp collisions} 
\label{sec_bbl}
An interesting feature of central $pp$ collisions at high energies
is that large--$x$ partons $(x \geq 0.01)$ in one nucleon pass through 
a strong gluon field in the other nucleon. This field can become so strong
that the interaction of the parton with the other nucleon approaches
the Black Body Limit (BBL), in which the probability for inelastic
scattering becomes unity, and the cross section becomes comparable
to the transverse size of the strong gluon field. This phenomenon
would have dramatic consequences for particle production in the 
forward region (dilepton production, hadron multiplicities and 
transverse momentum distributions), which have become the subject of
intensive theoretical investigation. In this section we want to
quantify the proximity to the BBL for $pp$ collisions at LHC energies.
Specifically, we want to show how a trigger on hard dijet 
(or multi--jet) production, which reduces the effective 
impact parameters in $pp$ collisions, greatly increases the
region (in the momentum fraction, $x$, and the virtuality, $Q^2$)
in which partons experience interactions close to
the BBL. For a recent review of approaches to ``taming'' 
the growth of parton densities at small $x$, based on the impact 
parameter eikonal approximation and the leading--log$x$ approximation, 
see Refs.\cite{McLerran:2003yx}.

To simplify the discussion, we consider instead of the scattering
of a colored parton the scattering of a small color--singlet dipole off the 
other nucleon. This is in the spirit of the dipole picture of high--energy 
scattering of Mueller \cite{Mueller:1994jq}.
The distribution of the inelastic and elastic cross sections, 
$\sigma^{dp}_{\text{in}}$ and
$\sigma^{dp}_{\text{el}}$, over the dipole--proton impact parameter, $\rho$, 
can be expressed in terms of the profile function of the dipole--nucleon 
elastic scattering amplitude, $\Gamma^{dp} (s, \rho )$
(\textit{cf.}\ Section~\ref{sec_inelastic}):
\be
\sigma^{dp}_{\text{in, el}} (s) \;\; = \;\; 
\int d^2 \rho \; \sigma^{dp}_{\text{in, el}} (s, \rho ),
\ee
with
\bea
\sigma^{dp}_{\text{el}} (s, \rho ) &=& |\Gamma^{dp} (s, \rho )|^2 , 
\label{sigma_el_Gamma} \\[2ex]
\sigma^{dp}_{\text{in}} (s, \rho ) &=& 
2 \; \text{Re}\; \Gamma^{dp} (s, \rho ) - |\Gamma^{dp} (s, \rho )|^2 .
\label{sigma_inel_Gamma}
\eea
The functions $\sigma^{dp}_{\text{in, el}} (s, \rho )$ are dimensionless and
can be interpreted as the ``cross sections per unit transverse area''.
If the target proton were a ``black'' disk of radius $R$, the probability
for a dipole hitting the target to undergo inelastic scattering would
be unity, and one would have $\sigma^{dp}_{\text{in}} (s, \rho ) = 1$ for 
$\rho < R$.
This would imply
\be
\Gamma^{dp} (s, \rho ) \;\; = \;\; 1 
\hspace{3em} \text{for} \;\;\; \rho < R.
\ee
From Eq.~(\ref{sigma_el_Gamma}) it follows that in this case 
the elastic cross section would also be unity
\be
\sigma^{dp}_{\text{el}} (s, \rho ) \;\; = \;\; 1
\hspace{3em} \text{for} \;\;\; \rho < R,
\ee
and thus the total (elastic plus inelastic) cross section would be 
twice the inelastic one. For realistic interactions, one may say
that the proton becomes black if $\Gamma^{dp} (s, \rho )$ 
approaches unity in a certain region of impact parameters.
In the following we shall use this criterion to quantify the
approach of dipole--nucleon interactions to the BBL. A similar
approach was used in Ref.~\cite{Rogers:2003vi} to study
the proximity of $\gamma^\ast N$ scattering to the BBL.

%
%
\begin{figure*}
\begin{tabular}{cc}
\includegraphics[width=8cm,height=8cm]{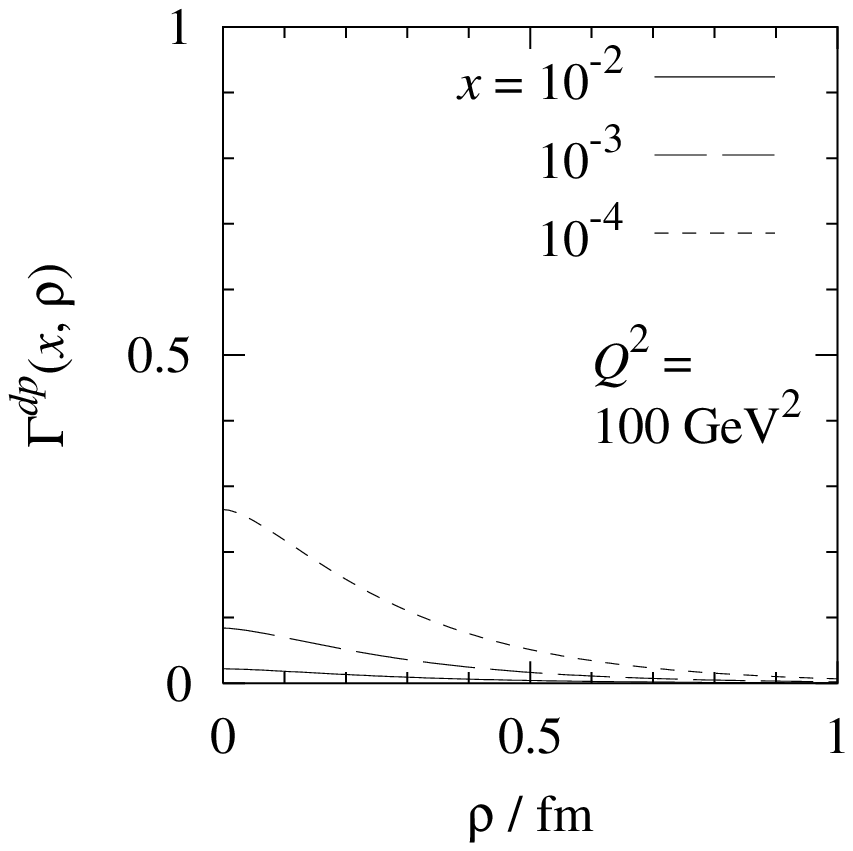}
& 
\includegraphics[width=8cm,height=8cm]{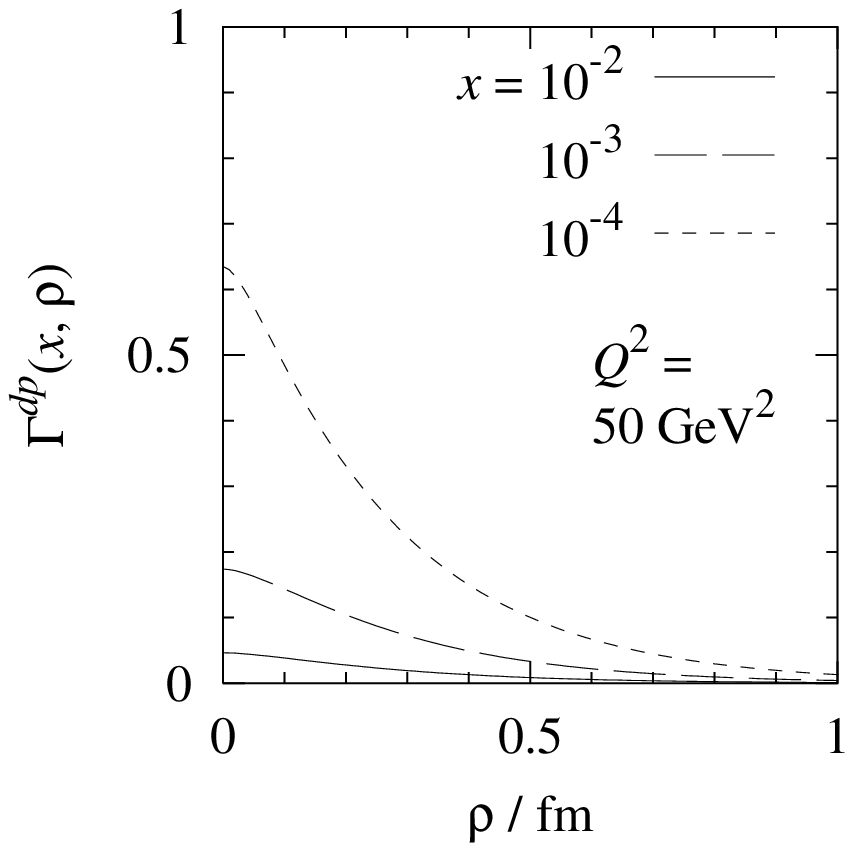}
\\
\includegraphics[width=8cm,height=8cm]{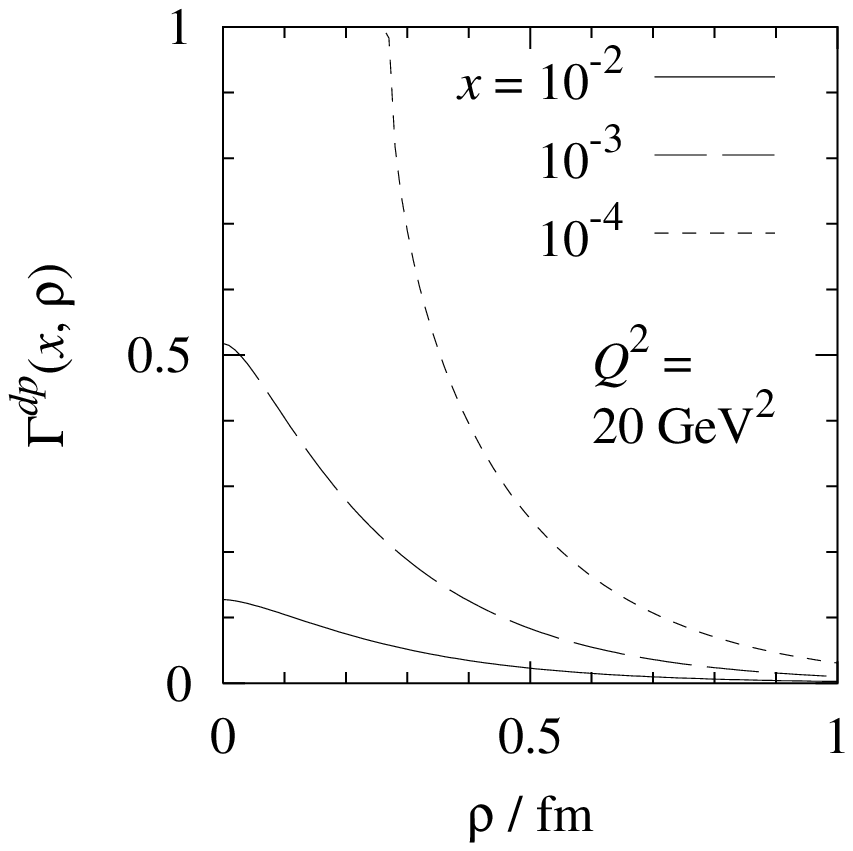}
&
\end{tabular}
\caption[]{The profile function for elastic 
dipole--nucleon scattering, $\Gamma^{dp} (x, \rho )$,
obtained from Eq.~(\ref{Gamma_from_sigma_in})
with the inelastic cross section given by the leading--twist expression,
Eq.~(\ref{sigma_LT}). Shown are the results for an $88$ dipole with
$Q^2 = 100 \, \text{GeV}^2$ (upper left panel), $50 \, \text{GeV}^2$
(upper right panel), and $20 \, \text{GeV}^2$ (lower left panel), 
for various values of x.}
\label{fig_Gamma_rho}
\end{figure*}
For small dipole sizes the cross section for inelastic 
dipole--nucleon scattering at fixed impact parameter $b$
is given by the leading--twist perturbative QCD 
expression \cite{Blaettel:rd,Frankfurt:it}
\be
\sigma^{dp}_{\text{in, LT}} (s, \rho ) \;\; = \;\;
\frac{C \pi^2}{3} \; d^2 \; \alpha_s (Q^2) \; x' \, g(x', \rho ; Q^2) .
\label{sigma_LT}
\ee
Here $C$ is a color factor, 
\be
C \;\; = \;\; 
\left\{
\begin{array}{ll}
1, & 3\bar 3 \; \text{dipole} \\[2ex]
9/4, & 88 \; \text{dipole}
\end{array} ,
\right.
\ee
$d$ is the dipole size, 
$\alpha_s (Q^2)$ the leading--order QCD running coupling, 
\bea
\alpha_s (Q^2) &=& \frac{4\pi}{\beta_0 \; 
\ln (Q^2 / \Lambda_{\text{QCD}}^2)} ,
\\
\beta_0 &\equiv& \frac{11}{3} N_c - \frac{2}{3} N_f , 
\eea
($\Lambda_{\text{QCD}}$ is the scale parameter), and $g(x, \rho ; Q^2)$ the 
impact parameter--dependent leading twist gluon density in the nucleon, 
\be
g(x', \rho ; Q^2) \;\; \equiv \;\; g(x', Q^2) \; F_g (x', \rho ; Q^2 ) ,
\ee
where $g(x', Q^2)$ is the usual (total) gluon density and 
$F_g (x', \rho ; Q^2 )$ the normalized $\rho$--profile, 
Eq.~(\ref{rhoprof_def}). The scale parameter $Q^2$ in $\alpha_s$ and
the gluon density is related to the dipole size by
\be
Q^2 \;\; = \;\; \frac{\lambda}{d^2} , 
\label{Q2_from_d2}
\ee
where $\lambda$ is a dimensionless parameter whose value is to be 
determined from phenomenological considerations. A value of 
$\lambda = 9$ sets $d$ equal to the average dipole size
contributing to the longitudinal photon--nucleon cross section
at large $Q^2$; we shall use this value of $\lambda$ in the following.
The gluon momentum fraction probed in inelastic 
dipole--nucleon scattering is determined by the invariant 
mass squared of the produced system, $M^2$,
\be
x' \;\; = \;\; \frac{M^2 + Q^2}{s} .
\ee
Here we have in mind generic inelastic production with $M^2 \sim Q^2$,
so we take
\be
x' \;\; \approx \;\; \frac{2 Q^2}{s} \;\; = \;\; 2 x 
\ee
as the argument for the gluon distribution. In the following we shall
regard the cross section \textit{viz.}\ the profile function as
a function of $x = Q^2 / s$ and the dipole--proton impact parameter, $\rho$.

Comparing Eqs.~(\ref{sigma_inel_Gamma}) and (\ref{sigma_LT}) we 
obtain an equation for the profile function $\Gamma$ 
(regarded now as a function of $x$ and $\rho$) corresponding to 
the leading--twist QCD result. If we assume the 
elastic amplitude to be imaginary, \textit{i.e.}, $\Gamma$ to be 
real, we obtain
\be
2 \, \Gamma^{dp} - (\Gamma^{dp})^2
\;\; = \;\; \sigma^{dp}_{\text{in, LT}} .
\ee
The relevant solution of this quadratic equation is
\be
\Gamma^{dp} (x, \rho ) \;\; = \;\; 1 - 
\sqrt{1 - \sigma^{dp}_{\text{in, LT}}(x, \rho )} .
\label{Gamma_from_sigma_in}
\ee
Note that this solution implies that $\Gamma^{dp} \rightarrow 0$,
and thus $\sigma_{\text{el}} \rightarrow 0$ 
[\textit{cf.}\ Eq.~(\ref{sigma_el_Gamma})],
for $\sigma_{\text{in, LT}} \rightarrow 0$, \textit{i.e.}, in the limit of 
small dipole size. For the corresponding solution with the ``+'' sign
in front of the square root the elastic cross section would tend to 
a constant value for zero dipole size. 

\begin{figure}
\includegraphics[width=8cm,height=8cm]{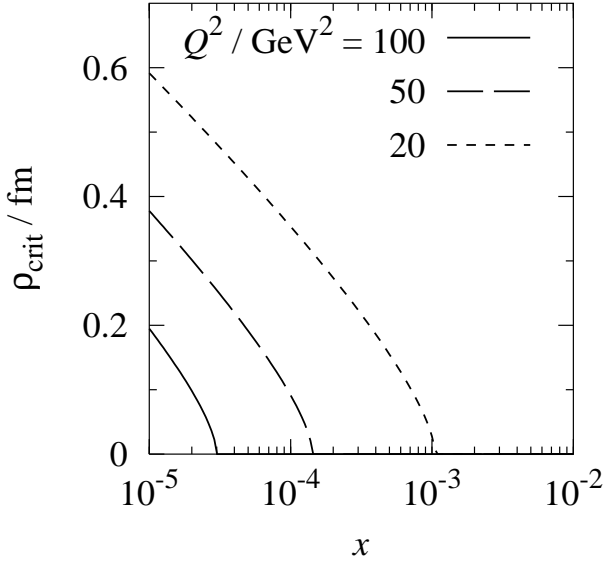}
\caption[]{The critical value of impact parameter, $\rho_{\text{crit}}$,
for which the profile function of elastic dipole--nucleon scattering,
$\Gamma^{dp} (x, \rho )$, exceeds the value 
$\Gamma^{dp}_{\text{crit}} = 0.5$,
\textit{cf.}\ Eq.~(\ref{rho_crit}). Here $\rho_{\text{crit}}$ is shown as 
a function of $x$. A value of $\rho_{\text{crit}} = 0$ implies that 
$\Gamma^{dp} (x, \rho ) < \Gamma^{dp}_{\text{crit}}$ 
for all values of $\rho$. Shown are the results for an $88$ dipole;
the $Q^2$--values are the same as in Fig.~\ref{fig_Gamma_rho}.}
\label{fig_rhoc}
\end{figure}

We evaluate the leading--twist expression for the inelastic 
$88$ dipole--nucleon cross section (\ref{sigma_LT}) with the
model for the $\rho$--dependent gluon distribution described
in Section~\ref{sec_hard} based on the dipole form factor 
with $x$-- and $Q^2$--dependent mass parameter.
Fig.~\ref{fig_Gamma_rho} shows the profile function 
$\Gamma^{dp} (x, \rho )$ obtained from Eq.~(\ref{Gamma_from_sigma_in}) 
for values $x = 10^{-2}, 10^{-3}$ and $10^{-4}$, and 
$Q^2 = 100, 50$ and $20\, \text{GeV}^2$;
the dipole size in each case is determined as $d^2 = \lambda / Q^2$
with $\lambda = 9$. The results clearly show the approach to the
BBL for decreasing $x$ (as a result of the growth of the gluon density)
and/or decreasing $Q^2$ (as a result of the increasing dipole size).
Note that the results for the profile function shown here are physically 
meaningful only in the region of impact parameters where 
$\Gamma^{dp}$ is significantly less than unity; 
for $\Gamma^{dp} \sim 1$ the simple leading--twist formula 
for the inelastic dipole--nucleon cross section, Eq.~(\ref{sigma_LT}),
breaks down. In other words, the figures should be read as indicating 
the \textit{region in impact parameter} for which the cross section 
approaches the BBL, rather than \textit{how} it is approached once the
interaction is sizable. An objective measure of the size of this region
is the value of $\rho$ at which $\Gamma^{dp}$ exceeds a certain critical
value, $\Gamma_{\text{crit}}$,
\be
\Gamma^{dp} (x, \rho ) \;\; \geq \;\; \Gamma^{dp}_{\text{crit}}
\hspace{3em} \text{for} \hspace{3em} \rho \; < \; \rho_{\text{crit}}.
\label{rho_crit}
\ee
Fig.~\ref{fig_rhoc} shows $\rho_{\text{crit}}$ as a function of $x$,
and for the above values of $Q^2$, for $\Gamma^{dp}_{\text{crit}} = 0.5$. 
This value of $\Gamma^{dp}$ corresponds to a probability of having no 
inelastic interaction of only 0.25, \textit{cf.}\ 
Eq.~(\ref{sigma_inel_Gamma}), which is a reasonable guideline. 
In Fig.~\ref{fig_rhoc} a value of $\rho_{\text{crit}} = 0$ implies 
that $\Gamma^{dp} (x, \rho )$ is smaller than $\Gamma^{dp}_{\text{crit}}$ 
for all values of $\rho$; even in the center of the nucleon where the 
gluon density is maximum ($\rho = 0$). 
%
%
%

%
%
\begin{figure}
\begin{tabular}{c}
\includegraphics[width=8cm,height=8cm]{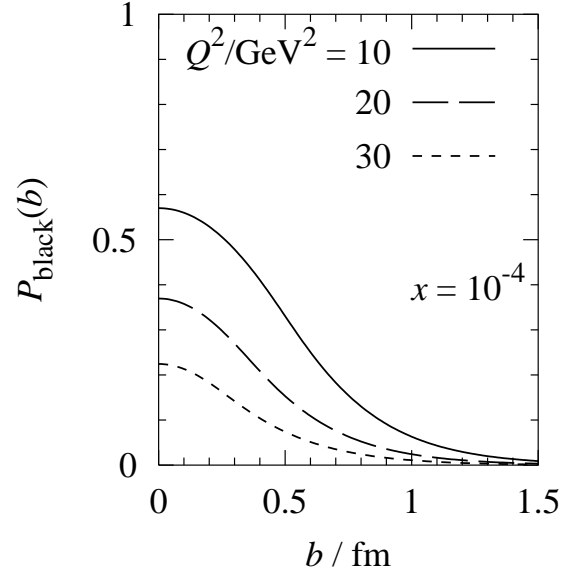}
\\
\includegraphics[width=8cm,height=8cm]{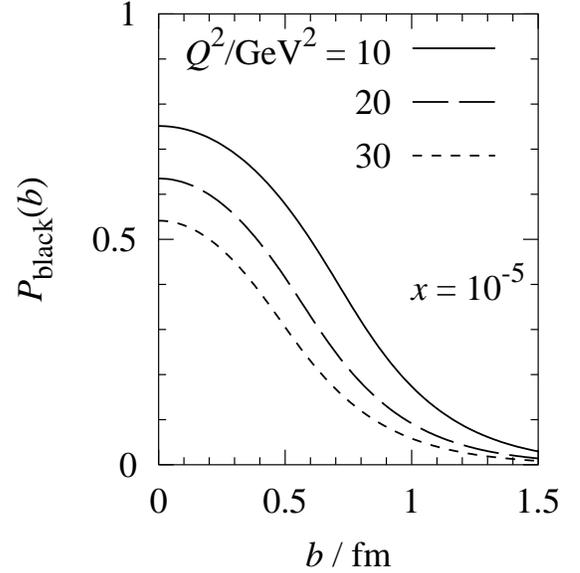}
\end{tabular}
\caption[]{The probability for partons in one proton with
given $x$ and $Q^2$ to interact with the other proton near the
BBL, as defined by Eq.~(\ref{P_black}), as a function of the 
impact parameter of the proton--proton collision, $b$. 
Shown are the results for $x = 10^{-4}$ (upper panel)
and $x = 10^{-5}$ (lower panel), for 
$Q^2 = 10, 20$ and $30 \, \text{GeV}^2$.}
\label{fig_P_black}
\end{figure}
We now turn to the question under which conditions
in proton--proton collisions a parton in one proton
will interact with the gluon field in the other proton near the BBL. 
The probability for this to happen depends on 
the transverse position of the parton relative to the center of its
parent proton, $\rho_1$, as well as the impact parameter of the 
proton--proton collision. An interesting measure is the total
probability for partons in one proton with given 
momentum fraction $x$ and virtuality $Q^2$ 
(but arbitrary transverse position)
to interact with the other proton near the BBL. 
It is given by the overlap integral of the 
normalized transverse spatial distribution of the partons
in their parent proton (shifted by the impact parameter vector, 
$\bm{b}$), with the characteristic function of the ``black'' region 
in the other proton, $\Theta (\rho_1 < \rho_{\text{crit}})$,
\bea
P_{\text{black}} (b) &\equiv&
\int d^2\rho_1 \; \Theta (\rho_1 < \rho_{\text{crit}}) \;
F_g (x, \rho_2 )
\nonumber \\
&& (\rho_2 \; \equiv \; |\bm{\rho}_1 - \bm{b}|).
\label{P_black}
\eea
This integral measures the fraction of the partons with given 
$x$ and $Q^2$ which hit the other proton in the ``black'' central region.
Here $\rho_{\text{crit}}$ depends on $x$ and $Q^2$, see above.
Generally, the probability $P_{\text{black}} (b)$ is maximum for central 
collisions $(b = 0)$ and decreases with increasing $b$. Note that if the 
target proton were a ``black'' disk of radius $R$, and the parent 
proton of the parton a disk of same size, $P_{\text{black}}$ would be unity 
at $b = 0$. Fig.~\ref{fig_P_black} shows $P_{\text{black}} (b)$ for partons 
with virtuality $Q^2 = 20 \, \text{GeV}^2$ and different values of $x$.

In actual hadron--hadron collisions, the parton which 
probes the gluon field in the other proton is resolved by a hard
collision with a parton in the other proton, resulting in hadron production,
see Fig.~\ref{fig_field}. Let $x_R$ be the momentum 
fraction of the resolving parton. Production of hadrons with transverse
momentum $p_\perp$ then resolves partons with
momentum fraction [\textit{cf.}\ Eq.~(\ref{x_resolved})]
\be
x \;\; = \;\; \frac{4 p_\perp^2}{x_R s}
\ee
and virtuality 
\be
Q^2 \;\; = \;\; 4 p_\perp^2 .
\ee
We are interested in partons with relatively large
$x_R$ ($\sim 10^{-1}$), which are able to resolve small--$x$
partons in the other proton whose interactions
are close to the BBL. Generally speaking, large $p_\perp$ 
select partons with large $x$ and large $Q^2$, 
for which interactions close to the BBL are unlikely.
Thus, the probability of interactions close to the BBL
decreases with increasing $p_\perp$. On the other hand, a 
certain minimum value of the parton virtuality, $Q^2$,
and thus of $p_\perp$, is required for the concept of
parton resolution by jet production to be applicable.
Thus, it is crucial to establish that there is a
``window'' in $p_\perp$ in which one is sensitive to BBL
effects while at the same time the partonic description
is still applicable. 

\begin{figure}
\includegraphics[width=8cm,height=8cm]{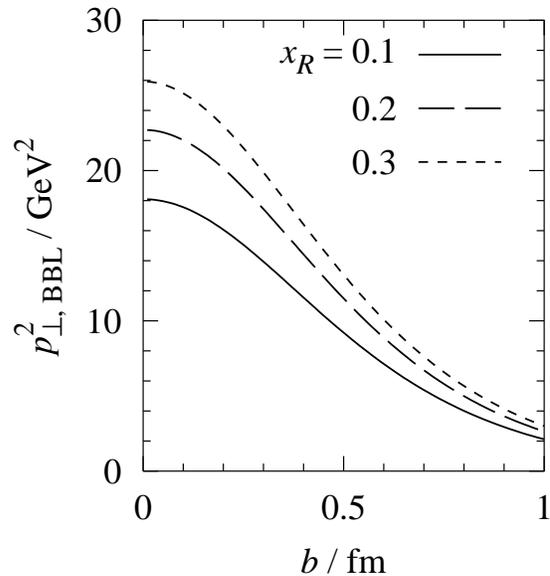}
\caption[]{The maximum value of the jet transverse momentum
squared, $p_{\perp , \text{BBL}}^2$, 
for which a small--$x$ parton in one proton, resolved in a
collision with a large--$x_R$ parton in the other proton,
interacts with the other proton close to the black--body limit 
(BBL), as a function of the impact parameter of the proton--proton 
collision, $b$. The criterion for proximity to the BBL 
is $P_{\text{black}} (b) > 1/2$, see Eq.~(\ref{P_black_crit}). 
Shown are the results for $x_R = 0.1, 0.2$ and $0.3$.}
\label{fig_pt}
\end{figure}

An important quantity is the maximum 
value of $p_\perp$ (for given $x_R$) for which the resolved 
parton sees the other proton as ``black''. 
We can estimate this maximum $p_\perp$ with the help of
the probability $P_{\text{black}}$ introduced in Eq.~(\ref{P_black}).
For given $x_R$, and given impact parameter $b$, we ask for the
maximum value of $p_\perp^2$ for which $P_{\text{black}}(b)$ exceeds
a certain critical value:
\be
P_{\text{black}} (b) \;\; > \;\; P_{\text{crit}}
\;\;\;\;\; \text{for} \;\;\;  p_\perp^2 \;\; < \;\; p_{\perp, \text{BBL}}^2 .
\label{P_black_crit}
\ee
Figure~\ref{fig_pt} shows $p_{\perp, \text{BBL}}^2$ corresponding
to the criterion that $P_{\text{black}} (b) > 0.5$, as a of the 
impact parameter of the proton--proton collision, $b$.
One sees that the maximum $p_\perp$ drops rapidly with increasing $b$.
[For $b \agt 1\, \text{fm}$ the maximum value of $p_\perp^2$
defined according to Eq.~(\ref{P_black_crit}) becomes $< 1 \, \text{GeV}^2$
and thus physically meaningless.] For central collision, 
$b < 0.5 \, \text{fm}$, the values of $p_{\perp, \text{BBL}}^2$ are larger
than $10 \, \text{GeV}^2$, so that the underlying leading--twist 
approximation is well justified. For such impact parameters it
is possible to explore the properties of the BBL using 
resolved hard partons as a well--defined probe.
%
%

Strictly speaking, the scattering of the small--$x$ parton in
one proton with the large--$x_R$ parton in the other proton 
occurs at finite  c.m. angles, 
and thus leads to a loss of light-cone fraction for the large--$x_R$
parton. This effect should be included in a more accurate treatment.

\begin{figure}
\includegraphics[width=8cm,height=8cm]{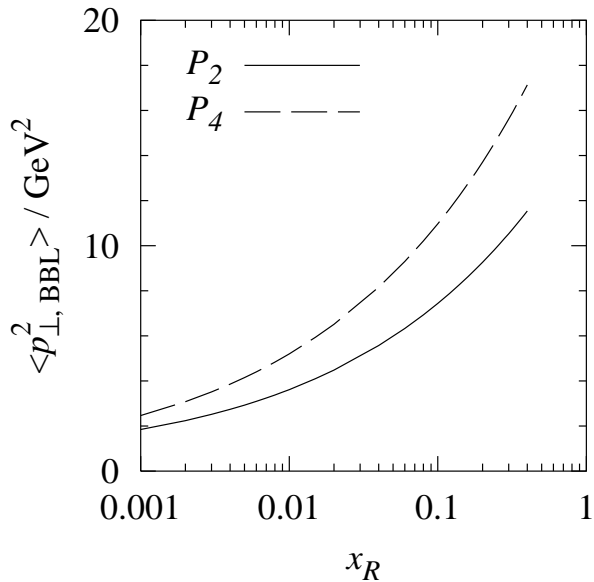}
\caption[]{The average value of $p_{\perp , \text{BBL}}^2$,
\textit{cf.}\ Fig.~\ref{fig_pt}, over impact parameters of the 
proton--proton collisions, Eq.~(\ref{ptav_def}), as a function of 
the resolving parton's momentum fraction, $x_R$.
Shown are the averages computed with the impact parameter distribution 
corresponding to the hard dijet trigger, $P_2$ (solid line), and the double 
dijet trigger, $P_4$ (dashed line), for $\sqrt{s} = 14000 \, \text{GeV}$
and dijet momentum $p_\perp = 25\, \text{GeV}$, 
\textit{cf.}\ Fig.~\ref{fig_P2_pt}.}
\label{fig_ptav}
\end{figure}

The above results clearly show that in order to explore the BBL 
with partons resolved in backward/forward jets as a probe 
one needs to limit the effective impact parameters in the $pp$ collisions. 
Here the idea of using hard dijet production at central rapidities as 
a ``centrality filter'', developed in Section~\ref{sec_trigger}, comes in.
Assuming that the simultaneous production of central and forward/backward 
jets can be described incoherently, \textit{i.e.}, in a probabilistic manner,
the probability for interactions near the BBL in events with hard dijet 
production is given by the average of the impact parameter--dependent
$p_{\perp, \text{BBL}}$ distribution of Fig.~\ref{fig_pt} with the effective 
$b$--distribution implied by the hard dijet trigger, $P_2 (b)$, see 
Section~\ref{sec_trigger}:
\be
\langle p_{\perp , \text{BBL}}^2  \rangle_2 \;\; \equiv \;\;
\int d^2 b \; p_{\perp , \text{BBL}}^2 (b) \; P_2 (b) .
\label{ptav_def}
\ee
A similar definition applies to the double dijet trigger with 
$b$--distribution $P_4 (b)$. 
For the LHC energy, $\sqrt{s} = 14000 \, \text{GeV}$, and
a dijet trigger with momentum $p_\perp = 25\, \text{GeV}$, 
\textit{cf.}\ Fig.~\ref{fig_pb}, the average values of 
$p_{\perp, \text{BBL}}^2$ obtained from Fig.~\ref{fig_pt} 
are shown in Fig.~\ref{fig_ptav} (solid line) as a function of the
resolving parton's momentum fraction, $x_R$. Also shown are the
corresponding averages with a double dijet trigger with the same 
$p_\perp$ (dashed line). One sees that the average values of
$p_{\perp, \text{BBL}}^2$ are all $\gg 1 \, \text{GeV}^2$, 
\textit{i.e.}, in the region where our assumption of resolved hard partons 
is well justified. What is equally important, the $b$--distributions 
implied by the hard multijet trigger suppress the contributions from 
large impact parameters, $b \gtrsim 1 \, \text{fm}$, where 
$p_{\perp , \text{BBL}}^2$ drops below $\sim 1 \, \text{GeV}^2$, 
meaning that the gluon density seen by the small--$x$ partons is so low 
that the BBL is never reached in the region where our approximations 
are justified. For the above dijet trigger, the fraction of events with $b$ 
so large that $p_{\perp , \text{BBL}}^2 (b) < 1 \, \text{GeV}^2$ is 
no larger that $\sim 10 \%$ for $x_R \geq 0.1$; 
for the double dijet trigger it drops to
$\sim 1 \%$. In contrast, for generic inelastic events without
the hard multijet trigger, the $b$--distributions of Fig.~\ref{fig_pb} 
show that a significant fraction of events would involve
$b$--values for which our perturbative treatment would not be
justified. In this sense, the possibility to control the effective 
impact parameters in the $pp$ collisions by way of the hard multijet 
trigger is more than just an enhancement of otherwise well--defined 
contributions --- it is crucial for the very applicability of the 
estimates presented here.
%
%

The above estimates are based on the interaction of color--octet ($88$) 
dipoles, corresponding to gluon partons, with the gluon field of the other 
proton. They can easily be extended to the case of color--triplet ($3\bar 3$) 
dipoles, corresponding to quark partons, see Eq.~(\ref{sigma_LT}) and after. 
In particular, one finds that in this case the values of 
$\langle p_{\perp , \text{BBL}}^2  \rangle$ are 
approximately $0.5$ times the values for gluons over the range of $x_R$
shown in Fig.~\ref{fig_ptav}, for both the dijet and the double 
dijet trigger.

It is worth noting that our estimate allows to avoid double counting 
due to multiple rescattering at the virtualities much smaller than the BBL. 
At the same time, it neglects an additional broadening due to the 
hard scattering at scales somewhat larger than BBL.
\section{Final state properties for central $pp$ collisions}
\label{sec_final}
We have seen that the probability for partons in one proton to 
interact with the other proton near the BBL is greatly increased 
by a trigger on hard dijet production, which acts as a 
centrality filter. We now outline the consequences of such interactions
close to the BBL for the final state of the hadronic collision.

Generally speaking, partons propagating through a strong gluon field 
get transverse momenta of the order of the maximum 
transverse momentum for which the interaction remains black,
$p_{\perp , \text{BBL}}$ \textit{cf.}\ the discussion in 
Refs.~\cite{Frankfurt:2001nt,Dumitru:2002qt,Dumitru:2002wd,Frankfurt:2002js}.
This value of transverse momentum is (approximately) related to the
maximum $Q^2$ at which the BBL is still valid as 
\be
p_{\perp , \text{BBL}} \;\; \approx \;\; Q_{\text{BBL}} /2 , 
\ee
assuming that in the hard interaction a system with mass of the order 
of $Q$ is produced. This relation can also be expressed as
\be
p_{\perp , \text{BBL}} \;\; \approx \;\; \frac{\sqrt{\lambda}}{2\, d}
\;\; = \;\; \frac{3}{2\, d},
\ee
where $d$ is the dipole size, \textit{cf.}\ Eq.~(\ref{Q2_from_d2}).
It is interesting that a similar numerical estimate follows
from the uncertainty relation: Regarding $d$ as conjugate to 
$p_{\perp}$ of the parton in the dipole one obtains 
$p_{\perp , \text{BBL}} \approx \pi/(2 d)$. Note that our
definition of the BBL scale, $Q_{\text{BBL}}$, is different
from the saturation scale $Q_s$ of the Color Glass Condensate
picture; nevertheless the two scales are numerically comparable.

We have seen that in the central collisions selected by the 
``centrality trigger'' the maximum transverse momenta squared, 
$p_{\perp, \text{BBL}}^2$, are $\gg 1 \, \text{GeV}^2$, 
\textit{cf.}\ Fig.~\ref{fig_ptav}.
This allows us to describe the hadronization of these partons
assuming independent fragmentation. In this approximation,
the differential cross section for the semi-inclusive production
of a given hadron $h$, characterized by its rapidity relative to 
the parent proton of the resolving parton, 
$\log z$, and transverse momentum , $k_\perp$,
is given by \cite{Dumitru:2002wd}
\bea
\lefteqn{z \frac{d\sigma^{pp\rightarrow hX}}{dz \, d^2 k_\perp} } &&
\nonumber \\
&=& \sum_{i = q, g} \int d^2 q_\perp \; d^2 l_\perp 
\int^1_z dx \; \frac{x}{z} \;
\nonumber \\
&\times& \delta^{(2)}\left( \bm{l}_\perp + (z/x) \bm{q}_\perp 
-\bm{k}_\perp \right)
\nonumber \\
&\times& 
f_{i/p}(x, Q^2_{\text{BBL}}) \; D_{i/p}(z/x, l_\perp , Q^2_{\text{BBL}}) \; 
c (q_\perp ).
\label{fragm}
\eea
Here $f_{i/p}(x, Q^2)$ is the leading--twist parton ($i =$ quark, gluon)
distribution in the proton, $c(q_\perp)$ the transverse momentum
distribution of the partons after passing through the strong gluon field,
which can be estimated in various models (see \textit{e.g.}\ 
Ref.~\cite{Dumitru:2002qt}),
and $D_{i/h}$ the parton fragmentation function.
Here we make the natural assumption, in line with the discussion above, 
that the small--$x$ partons are resolved at the scale 
$Q^2 = Q_{\text{BBL}}^2$, 
i.e.\ that the factorization scale is given by the maximal
virtuality at which the interaction is close to the BBL.
In Ref.~\cite{Dumitru:2002wd} a similar approach was applied to
particle production in central proton---nucleus collisions.
We stress that the approximation of independent fragmentation
is justified for sufficiently large values of $z$ (leading hadrons);
the relevant range in $z$ grows with the transverse momentum of the
produced hadrons. For small $z$ coherence effects need to be taken 
into account, see Ref.~\cite{Dokshitzer:nm} for a discussion.

In the BBL, estimates based on Eq.~(\ref{fragm}) should be regarded
as an upper bound for the spectrum of leading particles, since in the 
case of large--angle scattering a parton may actually convert into 
two high--$p_\perp$ partons with, on average, equal light cone
fractions, resulting in an even steeper drop of the leading hadron 
spectrum with $z$.  This is what happens in deep--inelastic scattering 
of a virtual photon in the BBL \cite{Frankfurt:2001nt}. 
In our case we expect somewhat smaller suppression
due to this effect, because large impact parameters contribute 
to the process (although with small probability), and in this
case much more forward particles would be produced. At the same time, 
the abovementioned trend of a steeper drop of the spectrum with $z$
should be much more pronounced for large--$k_\perp$ particles,
since in collisions at large impact parameters very few 
large--$k_\perp$ particles would be produced.

A detailed numerical investigation of hadron production 
in $pp$ collisions based on Eq.~(\ref{fragm}) is beyond the scope of the 
present paper. Here we only list some expectations for the 
qualitative properties of hadron production at large rapidities
which follow from independent fragmentation in the BBL:
\begin{itemize} 
\item 
The leading particle spectrum will be strongly suppressed
compared to interactions far from the BBL. The suppression will be 
especially pronounced for nucleons, so that for 
$z\agt 0.1$ the differential multiplicity of pions should exceed
that of nucleons.
\item The average transverse momenta of the leading particles 
will be $\agt 1\, \text{GeV}/c$.
\item There will typically be no correlation between the 
transverse momenta of leading hadrons, since they originate
from two different partons which have uncorrelated transverse momenta. 
Some correlations will remain, however, because two partons produced
in collisions of small--$x$ and large--$x$ partons may end up 
at similar rapidities.
\item 
For small impact parameters (which constitute a relatively small fraction 
of the total inelastic cross section but dominate in new particle
production) a large fraction of the events will have no particles with
$z\ge 0.02 \div 0.05$. This suppression will occur simultaneously in both 
fragmentation regions, corresponding to the emergence of long--range 
rapidity correlations between the fragmentation regions.
For studies of this feature of the central $pp$ collisions it would be
desirable to have good acceptance for both leading charged and neutral 
particles. This would allow one to measure the fraction of events 
without leading particles as a function of the centrality of the collision.
\item In the forward production of dimuons or dijets one expects
a broadening of the distribution over transverse momenta \cite{Gelis:2002fw}, 
as well as a weaker dependence of the dimuon production cross section 
on the dimuon mass for masses $\leq$ few GeV \cite{Frankfurt:2002js}.
\item The background for heavy particle or high--$p_\perp$ jet production
should contain a significant fraction of hadrons with transverse momenta
$p_\perp \sim p_{\perp, \text{BBL}}$, originating from fragmentation
of partons affected by the strong gluon field. The direction of 
the transverse momenta of these hadrons should be unrelated to the 
transverse momenta of the jets. This phenomenon will make it difficult 
to establish the direction of jets unless 
$p_\perp \; (\text{jet}) \gg p_{\perp, \text{BBL}}$.
\end{itemize}
In connection with particle production at large rapidities,
another interesting quantity to study will be the incident energy 
dependence of the leading particle $x_F$ multiplicity with the 
dijet ``centrality trigger'' for fixed values of $x_1$ and $x_2$.
In the absence of the trigger this would correspond to the study 
of usual Feynman scaling violations. In this case it is known that
different impact parameters contribute, and that leading particle 
production is predominantly a large impact parameter phenomenon,
which likely leads to a rather weak violation of Feyman scaling. 
On the other hand, in the ``conditional multiplicity'' with the 
centrality trigger small impact parameters give the dominant 
contribution, and thus the suppression of the forward spectrum should 
strongly increase with energy.

The proximity to the BBL in central $pp$ collisions will also lead to 
observable effects in particle production at small rapidities.
One expects a significant increase of the multiplicity at small 
rapidities, because interactions in the BBL will likely
lead to generation of large color charges in the fragmentation
regions. Also, as we discussed in Section~\ref{sec_trigger}, 
the production of multiple minijets will be strongly enhanced. 
Such an increase should in fact be present already at the Tevatron collider, 
in events with a trigger on two--jet or $Z^0$ production. We are aware of
only one study \cite{Field:2002vt} which investigated the correlation of the 
underlying event structure with the presence of such a trigger. 
An increase of the multiplicity at small rapidities was indeed observed.

The detailed modeling of the phenomena outlined in this section
will require building a Monte Carlo event generator, accounting for 
the proximity of the spectator interactions to the BBL and for the 
new pattern of flow of color excitations.
\section{Conclusions}
\label{sec_conclusions}
In this paper we have demonstrated that a trigger on hard dijet
production strongly reduces the effective impact parameters
in $pp$ collisions at LHC and, to some extent, at Tevatron energies.
The possibility to select central collisions with a reasonable
rate is of considerable practical importance.

We have argued that the structure of the final states in 
central $pp$ collisions at LHC energies will differ significantly 
from that of minimal bias events. The reason is that in the central
transverse region the interaction of hard partons with the gluon
field in the other proton approaches the unitarity (black--body) limit,
leading to an enhancement of transverse momenta and depletion of 
longitudinal momenta in hadron production at large rapidities.

The proposed centrality trigger offers new opportunities
for realistic studies of the physics of strong gluon fields in QCD.
On the experimental side, it singles out central events for which
the chances of reaching the BBL are maximized, and its signatures
in the final state can be clearly identified. On the theoretical side,
it quantitatively defines the region in transverse space in which 
one should expect deviations from DGLAP evolution due to unitarity effects.
This information should be incorporated in studies of non-linear 
QCD evolution within the renormalization group approach, 
see \textit{e.g.}\ Ref.~\cite{Rummukainen:2003ns} and references
therein. So far, such studies have assumed infinite extension
of the ``dense gluon medium'' in the transverse plane.
They have also neglected the contributions of large--$x$ 
partons to the parton densities at small $x$ due to 
$\log Q^2$--evolution, which are taken into account in our approach.

New heavy particles at LHC will be produced practically only 
in central collisions, as can be selected with the proposed 
hard dijet trigger. The fact that such collisions are also the
ones in which the interactions of hard partons reaches the BBL
poses new challenges for the analysis of such events. 
One needs to identify the signatures of new particle production
on top of the very specific modifications of the final state
implied by the BBL in the central region. This could imply
\textit{e.g.}\ changes in the definition of jets 
due to the enhanced pedestal of soft hadrons and enhanced production
of minijets, as well as changes in the cuts necessary to define
an isolated lepton. Furthermore, the sizable ``intrinsic'' $p_\perp$
of the partons acquired from interactions with the strong gluon
field may impose limits on the accuracy of the determination of the
masses of produced heavy particles.

There is an interesting connection of the $pp$ collisions at 
LHC energies discussed here with cosmic ray physics near the 
Greisen--Zatsepin--Kuzmin (GZK) cutoff, see Ref.~\cite{Nagano:ve} 
for a recent review. The density of gluons 
through which a proton propagates in a central $pp$ collisions at 
LHC energies is comparable to the typical density encountered 
by a proton near the GZK cutoff in collisions with air 
(\textit{i.e.}, light nuclei). Thus, the BBL effects on forward particle 
production described above could have an impact on the energy spectrum 
and composition of cosmic rays near the GZK cutoff.

Learning how hadron production depends on the impact parameter
of the $pp$ collision will allow to address also other questions of 
strong interaction dynamics, not primarily related to the BBL. 
For example, it is often argued that events with large hadron 
multiplicities (a factor of $\ge 2$ larger than average) 
at small rapidities are due to collisions at small impact parameters, 
in which the soft parton clouds of the colliding protons overlap 
much stronger than in average collisions. Once the correlation between 
the centrality of the collision and the suppression of the forward spectra 
\textit{etc.}\ has been established with the help of the dijet trigger, 
one would be able to check whether similar effects are present also 
in the high--multiplicity events, testing the hypothesis
that these are central collisions.

Finally, the comparison of our model estimate of the cross section
for double dijet production with the CDF data \cite{Abe:1997bp} 
indicates that there may be significant spatial correlations of 
partons in the transverse plane. This interesting phenomenon
should be investigated further. Not only is it important for 
possible extensions of the single dijet trigger to multiple dijet 
production --- it may also reveal interesting information about 
the structure of the nucleon at low scales, which generates the
parton distributions at the dijet production scale through 
$Q^2$--evolution.
\begin{acknowledgments}
The DGLAP evolution code used in the present study has been derived 
from a code obtained from J.~Kwiecinski several years ago. 
We are grateful to V.~Guzey and H.~Weigert for interesting theoretical
discussions, and to R.~D.~Field for discussions related to the CDF data. 
L.~F.\ acknowledges the hospitality of Penn State University. 
M.~S.\ thanks the Alexander--von--Humboldt Foundation for financial
support. C.~W.\ is supported by Deutsche Forschungsgemeinschaft.
This work has been supported by D.O.E.
\end{acknowledgments}

\begin{thebibliography}{99}
%
%
\bibitem{Gribov:fm}
V.~N.~Gribov,
Nucl.\ Phys.\  {\bf 22}, 249 (1961).
%
%
\bibitem{Block:1984ru}
M.~M.~Block and R.~N.~Cahn,
Rev.\ Mod.\ Phys.\  {\bf 57}, 563 (1985).
%
%
\bibitem{Frankfurt:2002ka}
L.~Frankfurt and M.~Strikman,
Phys.\ Rev.\ D {\bf 66}, 031502 (2002).
%
%
\bibitem{Abramowicz:1998ii}
H.~Abramowicz and A.~Caldwell,
Rev.\ Mod.\ Phys.\  {\bf 71}, 1275 (1999).
%
%
\bibitem{Frankfurt:2000ty}
L.~Frankfurt, V.~Guzey and M.~Strikman,
J.\ Phys.\ G {\bf 27}, R23 (2001).
%
%
\bibitem{Sjostrand:1987su}
T.~Sjostrand and M.~van Zijl,
Phys.\ Rev.\ D {\bf 36}, 2019 (1987).
%
%
\bibitem{Abe:1997bp}
F.~Abe \textit{et al.}  [CDF Collaboration],
Phys.\ Rev.\ Lett.\  {\bf 79}, 584 (1997);
Phys.\ Rev.\ D {\bf 56}, 3811 (1997).
%
%
\bibitem{Gribov:jg}
V.~N.~Gribov,
arXiv:hep-ph/0006158.
%
%
\bibitem{Islam:2002au}
M.~M.~Islam, R.~J.~Luddy and A.~V.~Prokudin,
Mod.\ Phys.\ Lett.\ A {\bf 18}, 743 (2003).
%
%
\bibitem{Kaidalov:2003vg}
A.~B.~Kaidalov, V.~A.~Khoze, A.~D.~Martin and M.~G.~Ryskin,
Acta Phys.\ Polon.\ B {\bf 34}, 3163 (2003).
%
%
\bibitem{Alberi:af}
G.~Alberi and G.~Goggi,
Phys.\ Rept.\  {\bf 74}, 1 (1981).
%
%
\bibitem{Strikman:2003gz}
M.~Strikman and C.~Weiss,
arXiv:hep-ph/0308191.
%
%
\bibitem{Burkardt:2002hr}
M.~Burkardt,
Int.\ J.\ Mod.\ Phys.\ A {\bf 18}, 173 (2003).
%
%
\bibitem{Pobylitsa:2002iu}
P.~V.~Pobylitsa,
Phys.\ Rev.\ D {\bf 66}, 094002 (2002).
%
%
\bibitem{Chekanov:2002xi}
S.~Chekanov \textit{et al.}  [ZEUS Collaboration],
Eur.\ Phys.\ J.\ C {\bf 24}, 345 (2002).
%
%
\bibitem{Kwiecinski:gc}
J.~Kwiecinski and D.~Strozik-Kotlorz,
Z.\ Phys.\ C {\bf 48}, 315 (1990).
%
%
\bibitem{Gluck:1998xa}
M.~Gluck, E.~Reya and A.~Vogt,
Eur.\ Phys.\ J.\ C {\bf 5}, 461 (1998).
%
%
\bibitem{Lippmaa:1997qb}
E.~Lippmaa \textit{et al.}  [FELIX Collaboration],
SLAC-R-638;
A.~Ageev \textit{et al.},
J.\ Phys.\ G {\bf 28}, R117 (2002).
%
%
\bibitem{Calucci:1999yz}
G.~Calucci and D.~Treleani,
Phys.\ Rev.\ D {\bf 60}, 054023 (1999).
%
%
\bibitem{Mueller:ey}
A.~H.~Mueller and H.~Navelet,
Nucl.\ Phys.\ B {\bf 282}, 727 (1987).
%
%
\bibitem{McLerran:2003yx}
L.~McLerran,
arXiv:hep-ph/0311028.
E.~Iancu and R.~Venugopalan,
arXiv:hep-ph/0303204.
%
%
\bibitem{Mueller:1994jq}
A.~H.~Mueller and B.~Patel,
Nucl.\ Phys.\ B {\bf 425}, 471 (1994).
%
%
\bibitem{Rogers:2003vi}
T.~Rogers, V.~Guzey, M.~Strikman and X.~Zu,
arXiv:hep-ph/0309099.
%
%
\bibitem{Blaettel:rd}
B.~Blaettel, G.~Baym, L.~L.~Frankfurt and M.~Strikman,
Phys.\ Rev.\ Lett.\  {\bf 70}, 896 (1993);
B.~Blaettel, G.~Baym, L.~L.~Frankfurt, H.~Heiselberg and M.~Strikman,
Phys.\ Rev.\ D {\bf 47}, 2761 (1993).
%
%
\bibitem{Frankfurt:it}
L.~Frankfurt, G.~A.~Miller and M.~Strikman,
Phys.\ Lett.\ B {\bf 304}, 1 (1993).
%
%
\bibitem{Frankfurt:2001nt}
L.~Frankfurt, V.~Guzey, M.~McDermott and M.~Strikman,
Phys.\ Rev.\ Lett.\  {\bf 87}, 192301 (2001).
%
%
\bibitem{Dumitru:2002qt}
A.~Dumitru and J.~Jalilian-Marian,
Phys.\ Rev.\ Lett.\  {\bf 89}, 022301 (2002).
%
%
\bibitem{Dumitru:2002wd}
A.~Dumitru, L.~Gerland and M.~Strikman,
Phys.\ Rev.\ Lett.\  {\bf 90}, 092301 (2003).
%
%
\bibitem{Frankfurt:2002js}
L.~Frankfurt and M.~Strikman,
Phys.\ Rev.\ Lett.\  {\bf 91}, 022301 (2003).
%
%
\bibitem{Dokshitzer:nm}
Y.~L.~Dokshitzer, V.~A.~Khoze, S.~I.~Troian and A.~H.~Mueller,
Rev.\ Mod.\ Phys.\  {\bf 60}, 373 (1988).
%
%
\bibitem{Gelis:2002fw}
F.~Gelis and J.~Jalilian-Marian,
Phys.\ Rev.\ D {\bf 66}, 094014 (2002).
%
%
\bibitem{Field:2002vt}
R.~D.~Field  [CDF Collaboration],
in Proceedings of the APS/DPF/DPB Summer Study on the Future of Particle 
Physics (Snowmass 2001), ed. N.~Graf, eConf {\bf C010630}, P501 (2001)
[arXiv:hep-ph/0201192];
T.~Affolder \textit{et al.}  [CDF Collaboration],
Phys.\ Rev.\ D {\bf 65}, 092002 (2002).
R.~Field  [CDF Collaboration],
Int.\ J.\ Mod.\ Phys.\ A {\bf 16S1A }, 250 (2001).
%
%
\bibitem{Rummukainen:2003ns}
K.~Rummukainen and H.~Weigert,
arXiv:hep-ph/0309306.
%
%
\bibitem{Nagano:ve}
M.~Nagano and A.~A.~Watson, 
Rev.\ Mod.\ Phys.\ {\bf 72}, 689 (2000).
%
%
\end{thebibliography}
\end{document}